\documentclass[aps,letterpaper,showpacs,12pt,floatfix]{revtex4}

\usepackage{amsmath}
\usepackage{bm}
\usepackage{graphicx}

\begin{document}

\preprint{\vbox{\hbox{JLAB-THY-05-288}}}

\title{ Modeling quark-hadron duality in polarization observables}

\author{Sabine Jeschonnek$^{(1)}$ and J. W. Van Orden$^{(2,3)}$}

\affiliation{\small \sl (1) The Ohio State University, Physics
Department, Lima, OH 45804\\
(2) Jefferson Lab, 12000 Jefferson Ave, Newport News, VA 23606 \\
(3) Department of Physics, Old Dominion University, Norfolk, VA
23529}

\date{\today}

\begin{abstract}

We apply a model for the study of quark-hadron duality in
inclusive electron scattering to the calculation of spin
observables. The model is based on solving the Dirac equation
numerically for a scalar confining linear potential and a vector
color Coulomb potential. We qualitatively reproduce the features
of quark-hadron duality for all potentials considered, and discuss
the onset of scaling and duality for the responses, spin structure
functions, and polarization asymmetries. Duality may be applied to
gain access to kinematic regions which are hard to access in deep
inelastic scattering, namely for $x_{Bj} \to 1$, and we discuss
which observables are most suitable for this application of
duality.

\end{abstract}

\maketitle

\section{Introduction}

Quark-hadron duality is a fascinating phenomenon that was first
observed by Bloom and Gilman \cite{bgduality} more than 30 years
ago. It is receiving plenty of attention today  from both the
experimentalist
\cite{jlab,f1fldual,prdmom,meziani,emcdual,hermes,hermesg1,e01012,12gevwp}
and theorist
\cite{closeisgur,closewallym,closezhao,closezhaogpd,carlnew,carlnimay,donghe,dongli,morechinese,
pp,mpdirac,marka1,kiev, leyaouancped, simula,wallymprl,
adelaide,ijmvo,jvod2,dirac,elba,myhugs,
zoltanfranz,liuti,davidovsky,hofmann} communities, due to an
interest in duality itself, and due to the huge field of
experimental applications of duality in kinematic regions that are
very difficult to access without it.

The most straightforward definition of quark-hadron duality says
that any hadronic process can be described in terms of either a
quark and gluon picture, or in terms of a purely hadronic picture,
provided either calculation contains all Fock states. However, in
the former case, a full numerical solution of QCD is prohibitive
in most situations, despite the impressive progress of lattice
QCD, and in the latter case, a full hadronic solution, e.g.
employing an effective field theory, also is not feasible unless
the kinematic region is restricted to low energies and momenta.
Thus, this most general version of duality is not very useful, as
many interesting processes take place in a region that is neither
perturbative nor
very low energy. %improve the "low energy" formulation

There is a much more practical version of duality: in certain
kinematic regimes, properly averaged hadronic observables can be
described by a perturbative QCD (pQCD) calculation. This version
of duality is highly relevant as perturbative QCD  calculations
can be performed. Using duality, these pQCD calculations can then
be related to averaged data taken in the resonance region.
Quark-hadron duality has been observed experimentally in many
processes: it was discovered by Bloom and Gilman in inclusive,
inelastic electron scattering, it made its way into the textbooks
in $e^+ e^- \to hadrons$, was studied in the semileptonic decays
of heavy mesons \cite{richural,semilep,isgurwise}, is considered
in the analysis of heavy ion reactions \cite{ralf}, and forms the
basis for using QCD sum rules \cite{qcdsr}. In addition to the
"classical" examples and applications  of duality, duality ideas
are applied in new areas, too. For neutrino scattering, the beam
energies are not well known, and an averaging will thus take place
almost automatically. The application of duality is discussed for
several planned neutrino experiments, see e.g. \cite{minerva}, and
duality ideas have been applied in \cite{Horowitz:2004yf} to
nucleon/nuclear duality in neutrino scattering. There is also
interest in duality in parity violation experiments \cite{vipuli},
and with regard to generalized parton distributions
\cite{closezhaogpd,diehl}. A very local version of duality -
assuming that it holds for just one resonance -  has been used in
\cite{wallymprl,adelaide} to extract information on structure
functions at $x_{Bj} \to 1$ in the scaling limit from form factor
data. These ideas were also applied to neutrino-nucleon scattering
\cite{adelaide}. Duality ideas might also be useful for pion
photoproduction \cite{haiyan}. Duality is a major point in the 12
GeV upgrade of CEBAF at Jefferson Lab \cite{12gevwp}.

In this paper, we investigate duality in inclusive, inelastic
electron scattering. New experimental data from Jefferson Lab and
DESY have impressively confirmed that quark-hadron duality is
valid down to rather low four-momentum transfers, and for many
observables: duality in $F_2$ was confirmed to hold down to $Q^2
\approx 0.5~GeV^2$ \cite{jlab}, and very recently, the
longitudinal structure function $F_L$ and the purely transverse
$F_1$ were separated, and found to exhibit duality for $Q^2
> 1 GeV^2$ \cite{f1fldual}. Experimental evidence for duality in spin observables
has been reported for $A_1^p$ from Hermes \cite{hermes} and from
Jefferson Lab \cite{meziani} for the first moment of $g_1^n$.
While this is exciting all by itself, these data have inspired
experimentalists to apply duality to the extraction of information
on the deep inelastic region in kinematics that are not readily
accessible. Duality allows us to connect the perturbative regime
of quarks and gluons with the strongly non-perturbative resonance
regime. The earliest example discussed was the extraction of the
elastic nucleon form factor from the deep inelastic scaling curve
\cite{dgp}. In \cite{jiunrau}, higher twist contributions were
inferred from the resonance data.

\begin{figure}[ht]
\includegraphics[width=20pc,angle=270]{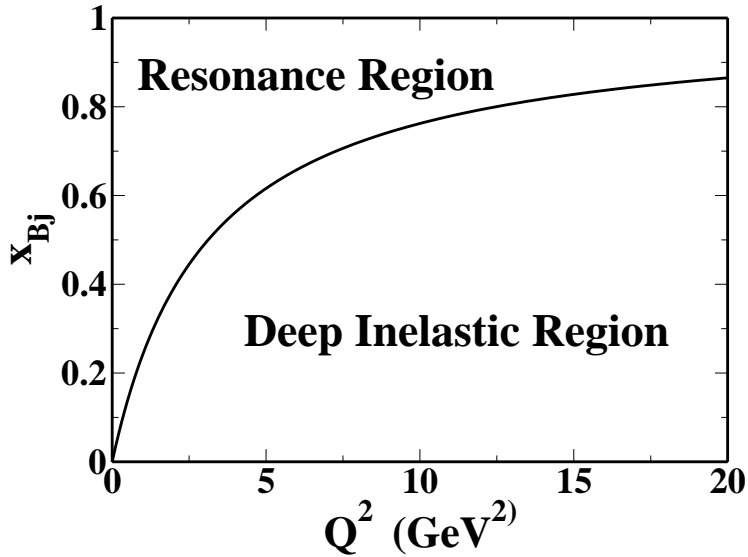}
\caption{Plot of the $x_{Bj}$ and $Q^2$ kinematic plane. This
kinematic plot shows the deep inelastic regime and the resonance
regime. The solid line corresponds to an invariant mass of the
final state of $W = 2~GeV$. Every point above the line lies in the
resonance region, $W < 2~GeV$, and every point below the line lies
in the deep inelastic region $W > 2~GeV$. The invariant mass of
the final state, $W$, is related to the four momentum transfer
$Q^2$ and Bjorken $x_{Bj}$ by $W^2 = M^2 + Q^2 (\frac{1}{x_{Bj}} -
1)$, where $M$ is the nucleon mass. } \label{kinplot}
\end{figure}

These two kinematical regions have traditionally been separated,
as it was believed that the physics of quarks and gluons had
little connection to the collective phenomena of resonances. At
low invariant masses $W$ of the final state in an $(e,e')$
reaction on the nucleon, one observes many resonance bumps in the
cross section, and $W < 2 ~GeV$ is traditionally referred to as
the resonance region. For higher invariant masses, $W > 2~GeV$,
the cross section becomes smooth and exhibits Bjorken scaling, and
this region is referred to as the deep inelastic region. Note that
this strict division of kinematics was introduced historically,
even though non-resonant processes are present for low $W$, and
 resonances with larger mass may contribute for high $W$.
The demarcation line of $W = 2~GeV$ is plotted in the $x_{Bj}$ -
$Q^2$ plane in Fig.~\ref{kinplot}.

One consequence of this traditional subdivision is that large
amounts of data with $W < 2 ~GeV$ were cut from deep inelastic
analyses of data, leading to a paucity of data at very high
Bjorken $x_{Bj}$. The region of large $x_{Bj}$, $x_{Bj} \to 1$, is
referred to as the (deep) valence region, and is the subject of
much interest. In particular, one would like to study the valence
quark spin distribution of the nucleon. This can be achieved by
measuring the polarization asymmetry $A_1$ of the proton and
neutron. For $x_{Bj} \to 1$, many, and widely different
predictions, exist for the polarization asymmetry of the neutron,
running the gamut from $0$, predicted in unbroken SU(6), to $1$,
predicted in pQCD, and everything in between, see
\cite{nathana1n,a1nprc} for good reviews of the situation. In the
chiral soliton model, there are even predictions of negative
values \cite{waka}.

The experimental data available for $A_1^n$ at high $x_{Bj}$ was
rather scarce and afflicted with very large error bars for $x_{Bj}
> 0.4$ before the advent of the recent Jefferson Lab data
\cite{a1datajlab}. They also extended only up to $x_{Bj} \approx
0.6$ \cite{smca1n,hermesa1n,slacg1,slaca1n}. This is due not only
to the fact that polarization experiments are always more
difficult to perform than unpolarized measurements, but mainly due
to the fact that in order to achieve $x_{Bj} \to 1$ for the deep
inelastic regime, one has to use very high four-momentum transfers
$Q^2$, see Fig.~\ref{kinplot}. This drastically reduces the cross
section, as the cross section is proportional to the Mott cross
section, which in turn is proportional to $1/Q^4$. Thus, accessing
one and the same $x_{Bj}$ in the resonance region and in the deep
inelastic regime leads to much lower count rates in the deep
inelastic regime. For example, a measurement at $x_{Bj} = 0.8$ can
be performed at $Q^2 = 2~ GeV^2$ in the resonance region, or at
$Q^2 \geq 15 ~GeV^2$ in the deep inelastic region. The count rate
in the deep inelastic region will be lowered by more than a factor
of 50 compared to the resonance region measurement. This makes
taking data in the deep inelastic regime for large $x_{Bj}$
extremely difficult. A recent Jefferson Lab experiment improved
the situation by measuring in the deep inelastic regime, up to
$x_{Bj} \geq 0.6$ with very reasonable error bars, thus decreasing
the uncertainty by an order of magnitude compared to older data
\cite{a1datajlab,a1nprc}. Still, the experimental exploration of
$A_1^n$ at really large $x_{Bj} \to 1$ has not been feasible yet,
thus making it the prime application for duality. Even with the
planned 12 GeV upgrade of Jefferson Lab, data in the deep
inelastic regime will be accessible only up to $x_{Bj} \approx
0.75$ \cite{12gevwp}.

As the kinematics plot shows, large values of $x_{Bj}$ can be
accessed at low $Q^2$ in the resonance regime. A measurement of
$A_1^n$ -  or any other observable of interest - in the resonance
region can then be averaged, and will yield the same information
as a direct measurement in the deep inelastic region, provided
that duality holds. Using this approach, $A_1^n$ could be measured
up to $x_{Bj} \approx 0.9$, by taking data in the Delta resonance
region \cite{nilanga}.

Another interesting application of duality that was recently
discussed is the application of duality to the EMC effect
\cite{emcdual}. Even though the EMC effect has been mapped out for
a large kinematic region, data at very high $x_{Bj}$ and for
lighter nuclei are scarce. The data base could be significantly
expanded by the application of duality to lepton scattering from
nuclei.

Before we can embark on this new experimental approach, we need a
good, solid understanding of how well duality holds, and where it
holds. Ideally, one would come up with a certain accuracy of the
duality procedure, which then could be quoted as a systematic
error for the extraction of $A_1^n$ or other observables in the
relevant region. Here, we need more theoretical input and
guidance. As we currently do not understand quark-hadron duality
from first principles, modeling is our best tool to obtain the
answers necessary for applying duality to extract $A_1^n$.

Currently, theorists  tackle this problem by modeling duality in
two different ways: one branch starts out from the
non-relativistic constituent quark model, with some relativistic
corrections, to describe duality
\cite{closeisgur,closewallym,donghe,dongli,morechinese}, the other
branch starts the modeling with a relativistic one-body equation
\cite{closezhao,closezhaogpd,pp,mpdirac,marka1,ijmvo,jvod2,dirac}.
The former branch makes contact with the phenomenology. It was
started by the pioneering work of Close and Isgur
\cite{closeisgur}, where the authors investigated how a summation
over the appropriate sets of nucleon resonances leads to parton
model results for the structure function ratios in the SU(6)
symmetric quark model. This work was recently expanded
\cite{closewallym} to include the effects of SU(6) spin-flavor
symmetry breaking. In \cite{donghe,dongli,morechinese}, the
authors considered the first five low-lying resonances, and
performed a careful analysis of the onset of duality for $F_2$ and
$g_1$. Our results belong to the latter branch. The goal of these
modeling efforts is obvious: to gain an understanding of
quark-hadron duality and the conditions under which it holds, by
capturing just the essential physical conditions of this rather
complex phenomenon. We imposed these basic requirements for a
model: we require a  relativistic description of confined valence
quarks, and we treat the hadrons in the infinitely narrow
resonance approximation.

This paper is the fourth in a series of papers, in which we have
modeled duality with increasing complexity. All models that we
have presented so far have reproduced the features of duality in a
qualitative manner. We started out with an all scalar model
\cite{ijmvo}, and gradually improved the model until all the
particles had proper spin \cite{dirac}.  In \cite{dirac}, we
focused on the model results for the unpolarized responses. For
the first time, we investigated the dependence of our results on
the type of potential we employed. In the present paper, we focus
on the spin observables: the responses $R_{T'}$ and $R_{TL'}$ that
are accessible only with spin, the spin structure functions $g_1$
and $g_2$, and the polarization asymmetries $A_1$ and $A_2$.

This paper is organized as follows: in Section \ref{secmodel}, we
briefly state the properties of the model, then, in Section
\ref{secformalism}, we discuss some of the formalism for polarized
inclusive electron scattering. In Section \ref{secresults} we
discuss our numerical results for the various spin observables for
the bound-free and the bound-bound transitions. We end with a
brief summary of our results and an outlook.

\section{The Model}
\label{secmodel}

We use the same model as in \cite{dirac}. For the convenience of
the reader, we present the key ingredients of our model here.

Our model consists of a constituent quark bound to an infinitely
heavy di-quark and is represented by the Dirac hamiltonian
\begin{equation}
\hat{H}=\bm{\alpha}\cdot\hat{\bm{p}}+\beta\left(m+V_s(r)\right)+V_v(r)\,,
\end{equation}
where the scalar potential is a linear confining potential given
by
\begin{equation}
V_s(r)=br,\qquad\qquad b=0.18{\rm GeV}^2\,.
\end{equation}
We have used the constituent quark mass in this paper, as our main
interest is the study of quark-hadron duality, which sets in at
rather low $Q^2$, experimentally $Q^2 \approx 0.5~GeV^2$ is
enough. In this kinematic region, the appropriate degree of
freedom is the constituent quark, which has acquired mass through
spontaneous chiral symmetry breaking. We have used a value for the
quark mass of $m = 258.46~MeV$ - obtained previously in a fit to
heavy mesons \cite{waw}. However, nothing hinges on using that
particular value: we changed our quark mass to $m = 10~MeV$, in
order to have a value reminiscent of a current quark mass, and
repeated our calculations. It turns out that, while scaling does
set in a little faster, there are no qualitative changes in the
results.

In our model, the vector potential is provided by a vector color
Coulomb potential. Calculations will be presented where the vector
color Coulomb potential is absent, that is $V_v(r)=0$, where the
vector potential is the simple static Coulomb potential
\begin{equation}
V_v (r) = V_c(r)=-\frac{4}{3}\frac{\alpha_s}{r}
\end{equation}
with $\alpha_s = 0.181$ and where the color Coulomb potential is
corrected to allow for the running coupling constant in a manner
similar to that used by Godfrey and Isgur \cite{godfreyisgur}. The
vector potential then has the form
\begin{equation}
V_v(r) = V_{cr}(r)=-\frac{4}{3r}\left(
\alpha_c\frac{1+e^-\frac{\rho_0}{\delta}}{1+e^\frac{\sqrt{b}r-\rho_0}{\delta}}
+\sum_{i=1}^2\alpha_i {\rm erf}(\gamma_i r)\right)
\end{equation}
where
\begin{eqnarray}
\alpha_c &=& 0.118 \nonumber\\
\rho_0&=& 0.04\nonumber\\
\delta&=& 0.01 \nonumber\\
\alpha_1&=& 0.239\nonumber\\
\alpha_2&=& 0.271 \nonumber\\
\gamma_1 &=& 0.746\ {\rm GeV}\nonumber\\
\gamma_2 &=&  5.40\ {\rm GeV} \,.
\end{eqnarray}
Note that we use different scalar and vector potentials, in
contrast to \cite{mpdirac,marka1}, where $V_s = V_v$ is used to
simplify the calculations.

We assume that only the light quark carries a charge, and we
choose unit charge for the light quark for simplicity.

\section{Spin Observables}
\label{secformalism}

In this section, we briefly review the formalism for calculating
responses for targets with arbitrary polarization axes, and
connect the definitions of the polarization asymmetries $A_1, A_2$
and the spin structure functions $g_1, g_2$ to the responses.

The hadronic tensor for targets with an arbitrary polarization
axis in $\hat s$ direction is:
\begin{equation}
W^{\mu \nu} = \sum_{m, m''} \omega^{\mu \nu}_{m m''} < \frac{1}{2}
m'' | \frac{1}{2} (1 + \vec \sigma \cdot \hat s) | \frac{1}{2} m>
\label{defhadtens}
\end{equation}
with
\begin{equation}
\omega^{\mu \nu}_{m m''} = \sum_{n' l' j' m'} < 1 0  \frac{1}{2}
m| J^{\mu \dagger} |n' l' j' m'> < n' l' j' m' | J^{\nu} | 1 0
\frac{1}{2} m''>
\end{equation}
where $J$ is the electromagnetic current operator. The ground
state's z-component of $j$ is denoted $m$ instead of $m_j$ for
brevity. Components of the hadronic tensor can be combined to give
\begin{eqnarray}
W_L & = & W^{00} \nonumber \\
W_T & = & W^{++} + W^{--} \nonumber \\
W_{TT} & = & 2 \Re(W^{+-}) \nonumber \\
W_{TL} & = & -2 \Re(W^{0+} - W^{0-}) \nonumber \\
W_{TL'} & = & -2 \Re(W^{0+} + W^{0-}) \nonumber \\
W_{T'} & = & W^{++} - W^{--}
\end{eqnarray}
The cross section for electron helicity $h$ and target
polarization axis $\hat s$ can be expressed in terms of the
hadronic tensor and the leptonic coefficients $v_K, K = L, T, TT,
TL, T', TL'$ as
\begin{equation}
\frac{d \sigma}{d E' d \Omega'} = \sigma_M \left ( v_L W_L + v_T
W_T + v_{TL} W_{TL} + v_{TT} W_{TT} + h \left [ v_{TL'} W_{TL'} +
v_{T'} W_{T'} \right ] \right ) \label{crossec}
\end{equation}
where $\sigma_{\rm Mott}$ is the Mott cross section, $\bm{q}$ is
the three-momentum transfer from the electron to the target, $\nu$
is the energy transfer and $Q^2=\bm{q}^2-\nu^2$, and $h$ denotes
the electron helicity. The leptonic coefficients are given by
\cite{donnellyras}
\begin{eqnarray}
v_L & = & \frac{Q^4}{\bm{q}^4} \nonumber \\
v_T & = & \frac{Q^2}{2\bm{q}^2}+\tan^2\frac{\theta}{2} \nonumber \\
v_{T'} & = &  \sqrt{\frac{Q^2}{ \bm{q}^2} +
\tan^2\frac{\theta}{2}} \tan
\frac{\theta}{2} \nonumber \\
v_{TL'} & = & - \frac{1}{\sqrt{2}} \frac{Q^2}{ \bm{q}^2} \tan
\frac{\theta}{2} \,.
\end{eqnarray}

%\begin{equation} \frac{d \sigma}{d E' d \Omega'} = \sigma_M \left
%( v_L W^{00} + v_T (W^{++} + W^{--}) + v_{TL} (-2 \Re(W^{0+} -
%W^{0-})) + v_{TT} (2 \Re(W^{+-})) + h \left [ v_{TL'} (-2
%\Re(W^{0+} + W^{0-})) + v_{T'} (W^{++} - W^{--}) \right ] \right )
%\end{equation}
For arbitrary target spin, all six combinations of the hadronic
tensor are non-zero. However, for spin $1/2$ targets, only four
%responses contribute: $R_L, R_T, R_{TL'}$ and $R_{T'}$.
combinations contribute: $L, T, TL'$ and $T'$. Inserting the
results for the current matrix elements, and exploiting selection
rules and symmetry relations between various current matrix
elements, one finds for our case
\begin{eqnarray}
W_L & = & \frac{1}{2} \sum_m \sum_{n' l' j'}  |< n' l' j' m | J^0
| 1 0 \frac{1}{2} m>|^2 (1 + s_z (-1)^{1/2 - m} ) \nonumber \\
W_T & = & \frac{1}{2} \sum_m \sum_{n' l' j'} (1 + s_z (-1)^{1/2 -
m} ) \nonumber \\
 & &  \hspace{2cm} \left [ |< n' l' j' m
+ 1 | J^+ | 1 0 \frac{1}{2} m>|^2  +  |< n' l' j' m - 1 | J^-
| 1 0 \frac{1}{2} m>|^2  \right ]\nonumber \\
W_{T'} & = & \frac{1}{2} \sum_m \sum_{n' l' j'} (1 + s_z (-1)^{1/2
- m} ) \nonumber \\
 & & \hspace{2cm} \left [ |< n' l' j' m + 1 | J^+ | 1 0 \frac{1}{2} m>|^2  -  |< n'
l' j' m - 1 | J^-
| 1 0 \frac{1}{2} m>|^2  \right ]\nonumber \\
W_{TT} & = & 0 \nonumber \\
W_{TL} & = & 0 \nonumber \\
W_{TL'} & = & - 2 s_x \sum_{n' l' j'} \Re ( <1 0 \frac{1}{2}
\frac{1}{2}|J_0^{\dagger} | n' l' j' \frac{1}{2}> <n' l' j'
\frac{1}{2}| J^+ | 1 0 \frac{1}{2} -\frac{1}{2}>)  \,.
\end{eqnarray}
Explicitly carrying out the summation over the initial spin $m$
for the $T'$ and $TL'$ combinations, and substituting for the
hadronic tensor combinations in the expression for the cross
section Eq.~(\ref{crossec}), we find:
\begin{eqnarray}
\frac{d \sigma}{d E' d \Omega'} & = & \sigma_M \bigg ( v_L
\frac{1}{2} \sum_m \sum_{n' l' j'} |J^0 (m)|^2  + v_T  \frac{1}{2}
\sum_m \sum_{n' l' j'} \left ( |J^+ (m)|^2 + |J^- (m)|^2 \right )
\nonumber \\
 +  & h &
 \left [- 2 s_x v_{TL'} \sum_{n' l' j'} \Re (J^{0, \dagger}
(\frac{1}{2}) J^+ (- \frac{1}{2}) ) + v_{T'} s_z  \sum_{n' l' j'}
\left (|J^+ (\frac{1}{2})|^2 - |J^- (-\frac{1}{2})|^2 \right
)\right ] \bigg )
\end{eqnarray}
where we abbreviated the current matrix elements as $J^{\mu} (m)$
for $< n' l' j' m | J^{\mu} | 1 0 \frac{1}{2} m>$. Now we can
relate this cross section to the definition of the polarization
observables. The polarization asymmetry $A_{||}$ is defined as the
ratio of the difference and sum of the cross sections for
longitudinally polarized electrons and target polarization
parallel or anti-parallel to the beam, see e.g.
\cite{weithom,jifil}:
\begin{equation}
A_{||} = \left [\frac{d \sigma^{\to \atop \leftarrow}}{d E' d
\Omega'} - \frac{d \sigma^{\to \atop \to}}{d E' d \Omega'} \right
] / \left [\frac{d \sigma^{\to \atop \leftarrow}}{d E' d \Omega'}
+ \frac{d \sigma^{\to \atop \to}}{d E' d \Omega'} \right ] \,.
\label{defapar}
\end{equation}
The upper superscript denotes the direction of the electron
polarization, the lower superscript indicates the direction of the
target polarization. The relevant cross sections are obtained by
using $h = 1$ in each case and $\hat s$ and $-\hat s$ for the
target polarizations parallel and antiparallel to the beam. Using
the conventional coordinate system, this means that the target is
polarized in the $x-z$ plane, see e.g. \cite{weithom,jifil}. In
this case, $\hat s = (\sin \alpha, 0, \cos \alpha)$, where
$\alpha$ is the angle between transferred momentum $\vec q$ and
beam momentum $\vec k$: $\cos \alpha = \frac{Q^2 + 2 E_{beam}
\nu}{2 E_{beam} q}$. Substituting and rearranging now yield:
\begin{eqnarray}
A_{||} & = & \displaystyle{\frac{v_{T'} s_z}{v_T + v_L
\frac{R_L}{R_T}}}
\nonumber \\
 & & \left ( \frac{- \sum_{n' l' j'} \left (|J^+ (\frac{1}{2})|^2 -
|J^- (-\frac{1}{2})|^2 \right )}{R_T} + 2 \, \frac{s_x}{s_z} \,
\frac{v_{TL'}}{v_{T'}} \, \,  \frac{\sum_{n' l' j'} \Re (J^{0,
\dagger} (\frac{1}{2}) J^+ (- \frac{1}{2})) }{R_T} \right )
 \label{apares}
\end{eqnarray}
Now, we compare this expression to the definition of the
polarization asymmetries $A_1, A_2$. For the convenience of the
reader, we quote the standard definitions \cite{weithom,jifil}:
\begin{equation}
A_{||} = D (A_1 + \eta A_2)
\end{equation}
with the depolarization factor $D$
\begin{equation}
D = \frac{1 - (1 - y) \epsilon}{1 + \epsilon R}
\end{equation}
where $y = \frac {\nu}{E_{beam}}$ and $R =
\frac{\sigma_L}{\sigma_T} = \frac{W_2}{W_1} (1 + \frac{\nu^2}{Q^2}
) - 1$, and
\begin{equation}
\eta = \frac{\epsilon \gamma y}{1 -  \epsilon (1 - y)}
\end{equation}
with  the magnitude of the virutal photon's longitudinal
polarization, $\epsilon = ( 1 + 2 \frac{q^2}{Q^2} \tan^2
\frac{\theta_e}{2} )^{-1}$, and $\gamma = \frac{2 M
x_{Bj}}{\sqrt{Q^2}}$. Note that the $y$ defined here is not the
$y$-variable used in the next section. Thus, we can read off the
expressions for the polarization asymmetries in terms of response
functions from Eq.~(\ref{apares}) as
\begin{eqnarray}
A_1 & = & - \frac{R_{T'}}{R_T} \nonumber \\
A_2 & = & - \frac{1}{\sqrt{2}} \frac{Q}{q} \frac{R_{TL'}}{R_T}
\end{eqnarray}
where we used the symmetry of the current matrix elements and the
definitions for the responses \cite{donnellyras}:
\begin{eqnarray}
R_L & = & \overline { \sum_{i,f}} |J^0|^2 \nonumber \\
R_T & = &  \overline {\sum_{i,f}} (|J^+|^2 + |J^-|^2) \nonumber \\
R_{T'} & = &  \overline {\sum_{i,f}} (|J^+|^2 - |J^-|^2) \nonumber \\
R_{TL'} & = & - 2  \overline {\sum_{i,f}} \Re( J^{0, \dagger} (J^+
+ J^-) )
\end{eqnarray}
The symbol $\displaystyle{\overline { \sum_{i,f}}}$ indicates the
average over initial states and the sum over final states. The
spin structure functions can be found as functions of the
responses using the relation between polarization asymmetries and
spin structure functions:
\begin{eqnarray}
A_1 & = & \frac{g_1 - \gamma^2 g_2}{F_1} \nonumber \\
A_2 & = & \gamma \frac{g_1 + g_2}{F_1}
\end{eqnarray}
with the unpolarized structure function $F_1 = M W_1 = \frac{1}{2}
M R_T$. We find:
\begin{eqnarray}
g_1 & = & - \frac{1}{2} M \frac{\nu^2}{q^2} (R_{T'} +
\frac{1}{\sqrt{2}} \frac{Q^2}{q \nu} R_{TL'}) \nonumber \\
g_2 & = & \frac{1}{2} M \frac{\nu^2}{q^2} (R_{T'} -
\frac{1}{\sqrt{2}} \frac{\nu}{q} R_{TL'})
\end{eqnarray}
Summarizing our expressions for the polarization observables in
terms of response functions, we have
\begin{eqnarray}
A_1 & = & - \frac{R_{T'}}{R_T} \nonumber \\
A_2 & = & - \frac{1}{\sqrt{2}} \frac{Q}{q} \frac{R_{TL'}}{R_T} \nonumber \\
g_1 & = & - \frac{1}{2} M \frac{\nu^2}{q^2} (R_{T'} +
\frac{1}{\sqrt{2}} \frac{Q^2}{q \nu} R_{TL'}) \nonumber \\
g_2 & = & \frac{1}{2} M \frac{\nu^2}{q^2} (R_{T'} -
\frac{1}{\sqrt{2}} \frac{\nu}{q} R_{TL'}) \,. \label{defpolobs}
\end{eqnarray}

For completeness, we also quote the expressions for the
unpolarized structure functions $W_1$ and $W_2$ in terms of the
responses:
\begin{equation}
W_1(Q^2,\nu)=\frac{1}{2}R_T(q,\nu)
\end{equation}
and
\begin{eqnarray}
W_2(Q^2,\nu)=\frac{Q^4}{\bm{q}^4}R_L(q,\nu)
+\frac{Q^2}{2\bm{q}^2}R_T(q,\nu)\,.
\end{eqnarray}

We focused on the longitudinal and transverse response functions
in our last paper \cite{dirac}, and we now focus on the true spin
observables, which we can now calculate, as quark spin is included
in our present model.

The Dirac wave functions and energy eigenvalues are obtained by
integrating the Dirac equation using the Runge-Kutta-Feldberg
technique and solutions are obtained for energies up to 12 GeV
with the radial quantum number of $n\cong 200$ and $|\kappa|\leq
70$.

In our model, we excite the bound quark from the ground state to
higher energy states, and do not allow it to decay. We refer to
this process as the bound-bound transition (previously, in
\cite{dirac}, we referred ot it as FSI). Thus, we do not include
any particle production in our model, and are strictly
quantum-mechanical in this sense. We do not have any gluons in our
model, either, which means that we do not encounter any radiative
corrections. Since the response functions consist of a sum of
delta functions, we choose to smear out the response functions by
folding with a narrow gaussian for purposes of visualization. The
smeared response functions are then given by
\begin{equation}
R_{K}(q,\nu)=\frac{1}{\sqrt{\pi}\epsilon}
\int_{-\infty}^{\infty}d\nu'\,e^{-\frac{(\nu-\nu')^2}{\epsilon^2}}
R_{K}^{unsmeared}(q,\nu')\,, \label{smear}
\end{equation}
where $K$ stands for $L, T, T'$ or $TL'$.

Before presenting numerical results, we would like to remind our
readers that, while the present model is more realistic than its
predecessors, its results should not be compared {\sl
quantitatively} to inclusive electron scattering from a nucleon.
Due to the assumption of an infinitely heavy antiquark (or
diquark) to which the light quark is bound, our calculation most
resembles inclusive electron scattering from a B-meson, which has
never been measured. The goal of our work is to gain a qualitative
understanding of duality, and the current simplification is no
impediment to this.

As discussed in \cite{dirac}, the current matrix elements
naturally are functions of the three-momentum transfer $q$.
Therefore, it is convenient to show our results for fixed
three-momentum transfer $q$ as a function of the $y$-scaling
variable

\begin{equation}
y=\sqrt{(\nu+E_0)^2-m^2}-q\,.\label{defy}
\end{equation}

The physics are not affected by presenting our results in this
fashion. In fact, later on we will explicitly compare our results
plotted for fixed $q$ and $y$ to our results plotted for fixed
four-momentum transfer $Q^2$ and the appropriate $x$ scaling
variable.

In the following, we will present analytic and numerical results
for the bound-free and bound-bound transitions, and we investigate
if duality holds or not. The conditions that need to be fulfilled
to see duality are: scaling of the bound-bound transition and
bound-free transition to the same scaling curve, and oscillation
of the bound-bound results at low momentum transfer around the
scaling curve. We are going to check if our model results
qualitatively reproduce the signature of duality as seen in the
electron scattering data.

In each case, we will investigate the approach to scaling for the
different observables, the scaling curves themselves, and the
behavior at low momentum transfers.

\section{Numerical Results}
\label{secresults}

Now, we turn to the numerical results of our model calculations.
Within our model, we calculate two different processes: the
bound-bound transition and  the bound-free transition of the light
quark. The bound-free transition (referred to as PWIA in
\cite{dirac}), is the analog to perturbative QCD. The bound quark
is knocked into the continuum by the absorption of the virtual
photon.

We discuss the bound-free transition first, employing just the
linear confining potential. We investigate the scaling results and
the onset of scaling for the bound-free transition. Next, in
Section \ref{secbb}, we show our results for the bound-bound
transition and the linear confining potential. There, we focus on
the onset of scaling and the low-$q$ duality. Then, we take a look
at the role of the ground state p-wave contribution in Section
\ref{secpwave}. Finally, we investigate the effect of employing a
static Coulomb and a running Coulomb potential in Section
\ref{secpotmodels}.

\subsection{y-scaling for the bound-free transition}
\label{secpwiayscal}

For the bound-free transition, we can reach arbitrarily high
energy transfers without any numerical complications, as we do not
have to solve for any high energy bound states in the final state.
Moreover, we can determine the analytic expressions for the
responses, and thus for all other observables, for $q \to \infty$
at fixed $y$. As the responses $R_L$ and $R_T$, which can be
accessed in unpolarized scattering, enter some spin observables,
we quote all four responses for the reader's convenience
\cite{dirac}:

\begin{eqnarray}
R_L(q,y)&=&\frac{1}{16\pi^2q}\int_{|y|}^{y+2q}dp p
\biggl\{\sqrt{(y+q)^2+m^2}n^0_v(p)+m\,n_s(p)\nonumber\\
&&\left.+ \frac{y^2+2qy+p^2}{2p}\,n^s_v(p)\right\}
\end{eqnarray}
\begin{eqnarray}
R_T(q,y)&=&\frac{1}{8\pi^2q}\int_{|y|}^{y+2q}dp p
\biggl\{\sqrt{(y+q)^2+m^2}n^0_v(p)-m\,n_s(p)\nonumber\\
&&\left.- \frac{y^2+2qy+2q^2-p^2}{2q}
\,\frac{y^2+2qy-p^2}{2pq}\,n^s_v(p)\right\}\,,
\end{eqnarray}
\begin{eqnarray}
R_{T'}(q,y)&=& - \frac{1}{8\pi^2q}\int_{|y|}^{y+2q}dp p \biggl\{
\left[ \left(\sqrt{(y+q)^2+m^2} + m \right) n_-(p)  -p   n^s_v(p)
\right] \left( \frac{(y^2 + 2 y q - p^2}{2 p q} \right)^2 \nonumber\\
&&\left.
+ \sqrt{(y+q)^2+m^2} \, n_s(p) - m \, n^0_v(p) - \frac{y^2 + 2 y q
-p^2}{2p} \, n^s_v(p)\right\}\,,
\end{eqnarray}
and
\begin{eqnarray}
R_{TL'}(q,y)&=&\frac{\sqrt{2}}{8\pi^2q}\int_{|y|}^{y+2q} dp p
\biggl\{ \frac{q}{2}  \left [ 1 - \left( \frac{(y^2 + 2 y q -
p^2}{2 p q} \right)^2 \right ] n_-(p) + q \, n_s(p)
\nonumber\\
&&\left.- \frac{y^2 + 2 y q - p^2}{2 p q} \big( m \, n^s_v(p) - p
\, n_s(p) \big) \right\}\,.
\end{eqnarray}

In the limit of large $q$, the bound-free response functions
become
\begin{eqnarray}
\lim_{q\rightarrow\infty}R_L(q,y)&=&\frac{1}{16\pi^2}\int_{|y|}^{\infty}dp
p \left\{n^0_v(p)+ \frac{y}{p}\,n^s_v(p)\right\}
\end{eqnarray}
%
%and
%
\begin{eqnarray}
\lim_{q\rightarrow\infty}R_T(q,y)&=&\frac{1}{8\pi^2}\int_{|y|}^{\infty}dp
p \left\{n^0_v(p)- \frac{y}{p}\,n^s_v(p)\right\}\,,
\end{eqnarray}
\begin{eqnarray}
\lim_{q\rightarrow\infty}R_{T'}(q,y)&=& -
\frac{1}{8\pi^2}\int_{|y|}^{\infty}dp p \left\{\frac{y^2}{p^2} \,
n_-(p) + n_s (p)- \frac{y}{p}\,n^s_v(p)\right\}\,,
\end{eqnarray}
and
\begin{eqnarray}
\lim_{q\rightarrow\infty}R_{TL'}(q,y)&=&\frac{\sqrt{2}}{8\pi^2}\int_{|y|}^{\infty}dp
p \left\{(1 -\frac{y^2}{p^2}) \, \frac{1}{2} \, n_-(p) + n_s (p)
\right\}\,.
\end{eqnarray}

These response functions therefore scale in $y$.

The vector and scalar momentum density distributions $n_v(p)$ and
$n_s(p)$ are defined in terms of the ground state wave function
\begin{equation}
\Psi_{10\frac{1}{2}m}(\bm{p})=\left(\begin{array}{c}
\psi^{(+)}_{10\frac{1}{2}}(p){\cal Y}_{0\frac{1}{2}}^m(\Omega_p)\\
\psi^{(-)}_{10\frac{1}{2}}(p){\cal Y}_{1\frac{1}{2}}^m(\Omega_p)
\end{array}\right)
\end{equation}
as
\begin{equation}
n_v(p)=\left(n_v^0(p), \frac{\bm{p}}{|\bm{p}|}n_v^s(p)\right)
\end{equation}
with
\begin{equation}
n_v^0(p)=\frac{1}{2\pi}\left({\psi^{(+)}_{10\frac{1}{2}}}^2(p)+
{\psi^{(-)}_{10\frac{1}{2}}}^2(p)\right)
\end{equation}
and
\begin{equation}
n_v^s(p)=\frac{1}{\pi}\psi^{(+)}_{10\frac{1}{2}}(p)
\psi^{(-)}_{10\frac{1}{2}}(p)\,;
\end{equation}
and
\begin{equation}
n_s(p)=\frac{1}{2\pi}\left({\psi^{(+)}_{10\frac{1}{2}}}^2(p)-
{\psi^{(-)}_{10\frac{1}{2}}}^2(p)\right)\,;
\end{equation}
moreover,
\begin{equation}
n_-(p)=\frac{1}{\pi}\left({\psi^{(-)}_{10\frac{1}{2}}}^2(p)
\right)\,.
\end{equation}

\begin{figure}[ht]
\includegraphics[width=20pc,angle=270]{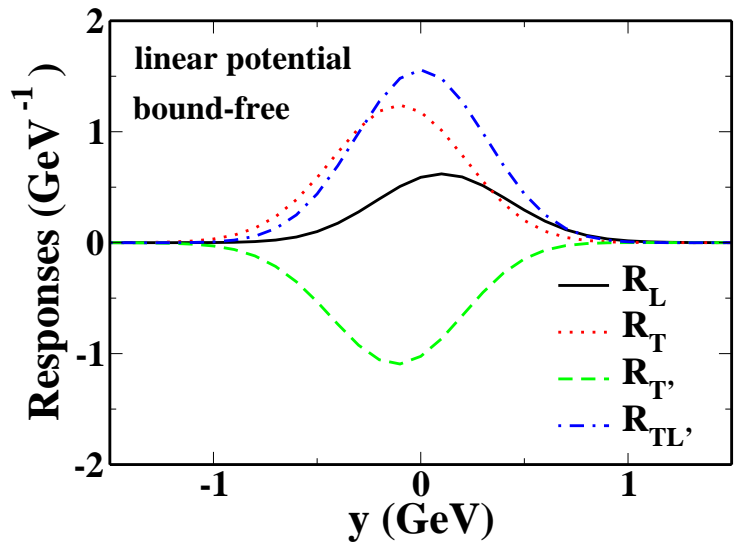}
\caption{The asymptotic behavior for $q \to \infty$ and fixed $y$
of the responses for the bound-free transition. We show $R_L$ (solid
line), $R_T$ (dotted line), $R_{T'}$ (dashed line), and $R_{TL'}$
(dash-dotted line).}
\label{figbfasymresp}
\end{figure}

The asymptotic responses for the bound-free transition are shown
in Fig. \ref{figbfasymresp}.  The two purely transverse responses,
$R_T$ and $R_{T'}$, have opposite signs but similar peak positions
at slightly negative $y$ values and peak heights. The longitudinal
response $R_L$ is the smallest of the four responses, and the only
one to peak at a slightly positive $y$. The
transverse-longitudinal interference response is the largest
response, and peaks at $y = 0$. This response is symmetric around
$y = 0$ as it contains only terms proportional to $y^0$ and $y^2$,
but no contributions linear in $y$.

\begin{figure}[ht]
\includegraphics[width=20pc,angle=270]{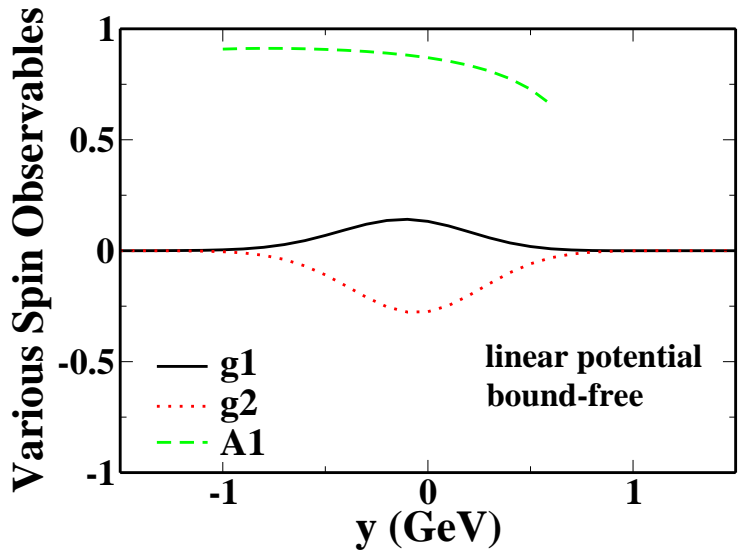}
\caption{The asymptotic behavior for $q \to \infty$ and fixed $y$
of the spin structure functions $g_1$ and $g_2$ and the polarization
asymmetry $A_1$ for the bound-free transition. We show $g_1$ (solid
line), $g_2$ (dotted line), and $A_1$ (dashed line). Note that $A_2$
vanishes in this limit.}
\label{figbfasymssf}
\end{figure}

The similar size and shape of $R_T$ and $R_{T'}$, together with
their opposite signs, leads to a value of the polarization
asymmetry $A_1 = - \frac{R_{T'}}{R_T}$ close to 1, see
Fig.~\ref{figbfasymssf}.  We show $A_1$ only for $y$ values for
which the responses have appreciable values, i.e. for $-1 < y <
0.6$. While we could calculate $A_1$ in a region where the
responses are tiny, the results would be meaningless. Our result
is close to the pQCD prediction of 1 for $A_1$ in the limit of $x
\to 1$. For large $Q^2$, large negative values of $y$ - much
bigger than $y = -1 GeV$ - correspond to $x = 1$.

The asymptotic values of the spin structure functions $g_1$ and
$g_2$ are also shown in Fig.~\ref{figbfasymssf}.  The spin
structure functions have peaks at small, negative $y$ values. They
have opposite signs. Of the two functions, $g_2$ has the slightly
larger peak value. Note that the asymptotic value of $A_2$ is
zero.

\subsection{Approach to scaling in the bound-free transition}

Now, after showing the asymptotic values, we will consider the
approach to scaling in the bound-free transition. Previously, we
have studied the approach to scaling of the unpolarized
longitudinal and transverse response functions \cite{dirac}.
There, we found that the onset of scaling is not influenced very
much by the spin of the target particle. The two response
functions $R_{T'}$ and $R_{TL'}$ are accessible only with
polarized beams and targets. It is interesting to check if they
scale just like their unpolarized counterparts.

\begin{figure}[ht]
\includegraphics[width=20pc,angle=270]{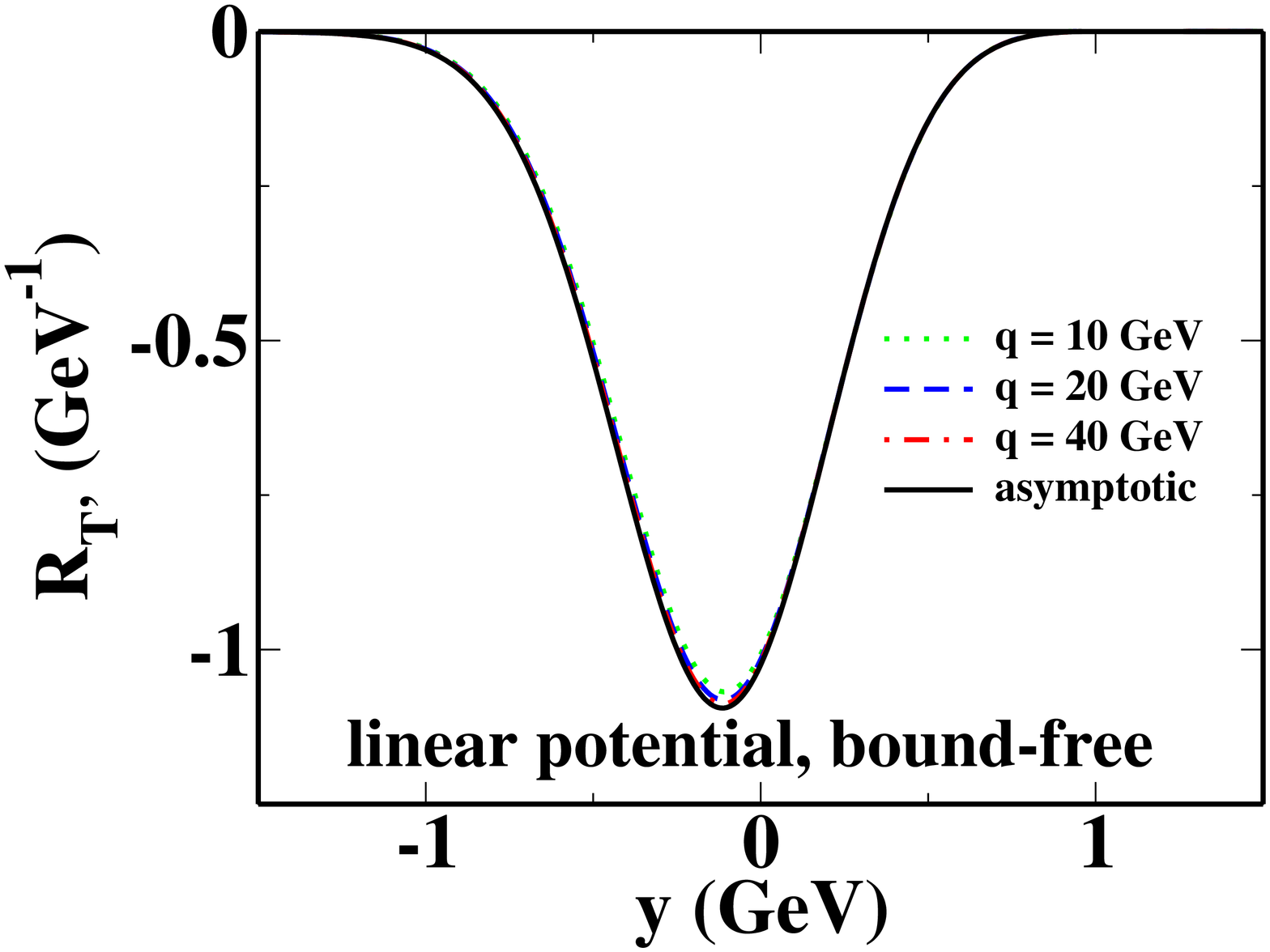}
\includegraphics[width=20pc,angle=270]{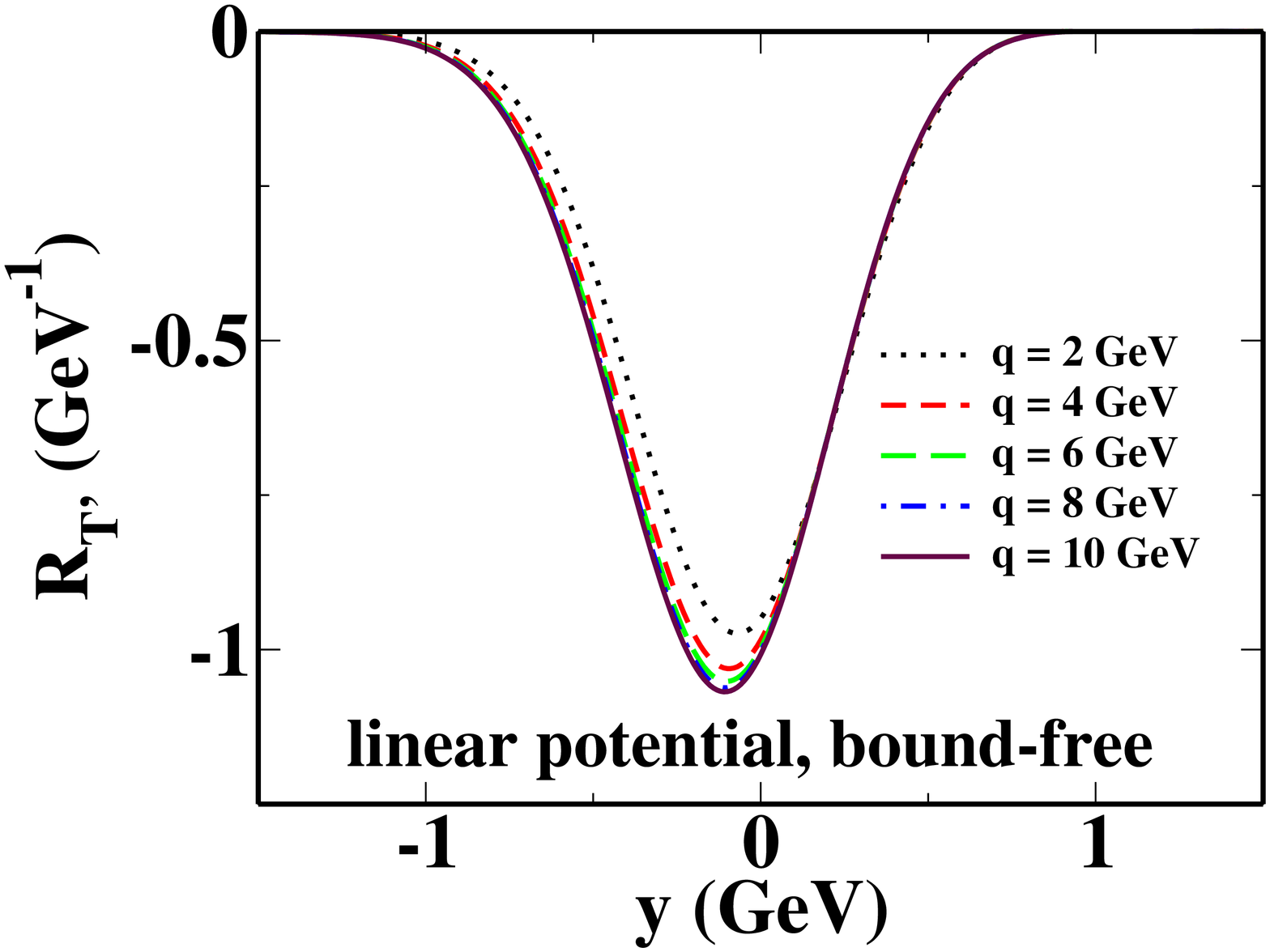}
\caption{The transverse-primed response function $R_{T'}$ is plotted
 versus $y$ for several low (bottom panel) and high (top panel) values
of the three-momentum transfer $q$. The results shown have been
calculated for the bound-free transition, using the linear potential.}
\label{figbfrtp}
\end{figure}

The scaling behavior of the transverse-primed response $R_{T'}$ for
the bound-free transition is shown in Fig.~\ref{figbfrtp}. The lower
panel shows the response for several $q$ values up to $10$~GeV. With
increasing momentum transfer, the peak of the response moves to more
negative values of $y$, and increases in height. The width of the
response also increases. The change in the response when going from $q
= 2 GeV$ to $q = 4 GeV$ is significant, and the difference between the
$q = 4 GeV$ and $q = 6 GeV$ results is noticeable. Increasing $q$
leads to very small changes, visible mainly at the peak and the large,
negative $y$ flank. The top panel, displaying the response for various
higher values of $q$ and the asymptotic value discussed above, shows
that the response has converged to the asymptotic value roughly at $q = 40
GeV$.

\begin{figure}[ht]
\includegraphics[width=20pc,angle=270]{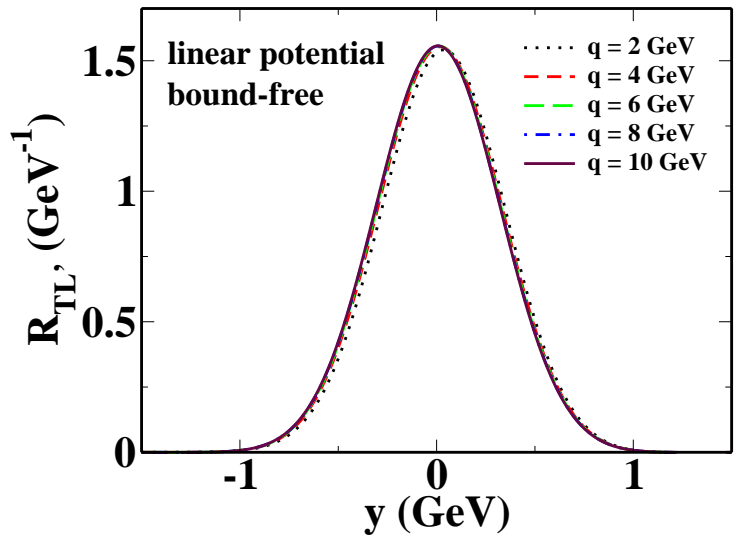}
\caption{The transverse-primed response function $R_{TL'}$ is plotted
 versus $y$ for several values of the three-momentum transfer $q$. The
results shown have been calculated for the bound-free transition,
using the linear potential.}
\label{figbfrtlp}
\end{figure}

The transverse-longitudinal primed response $R_{TL'}$ for the
bound-free transition is shown in Fig.~\ref{figbfrtlp}. The figure
shows the response for lower values of the three-momentum transfer
$q$. While there are small changes, amounting to a small shift of the
entire response towards the negative $y$ values, one can see that this
response scales much faster than $R_{T'}$. We do not include another
panel with $R_{TL'}$ calculated for higher $q$ values, as the curves
all coincide.

\begin{figure}[ht]
\includegraphics[width=20pc,angle=270]{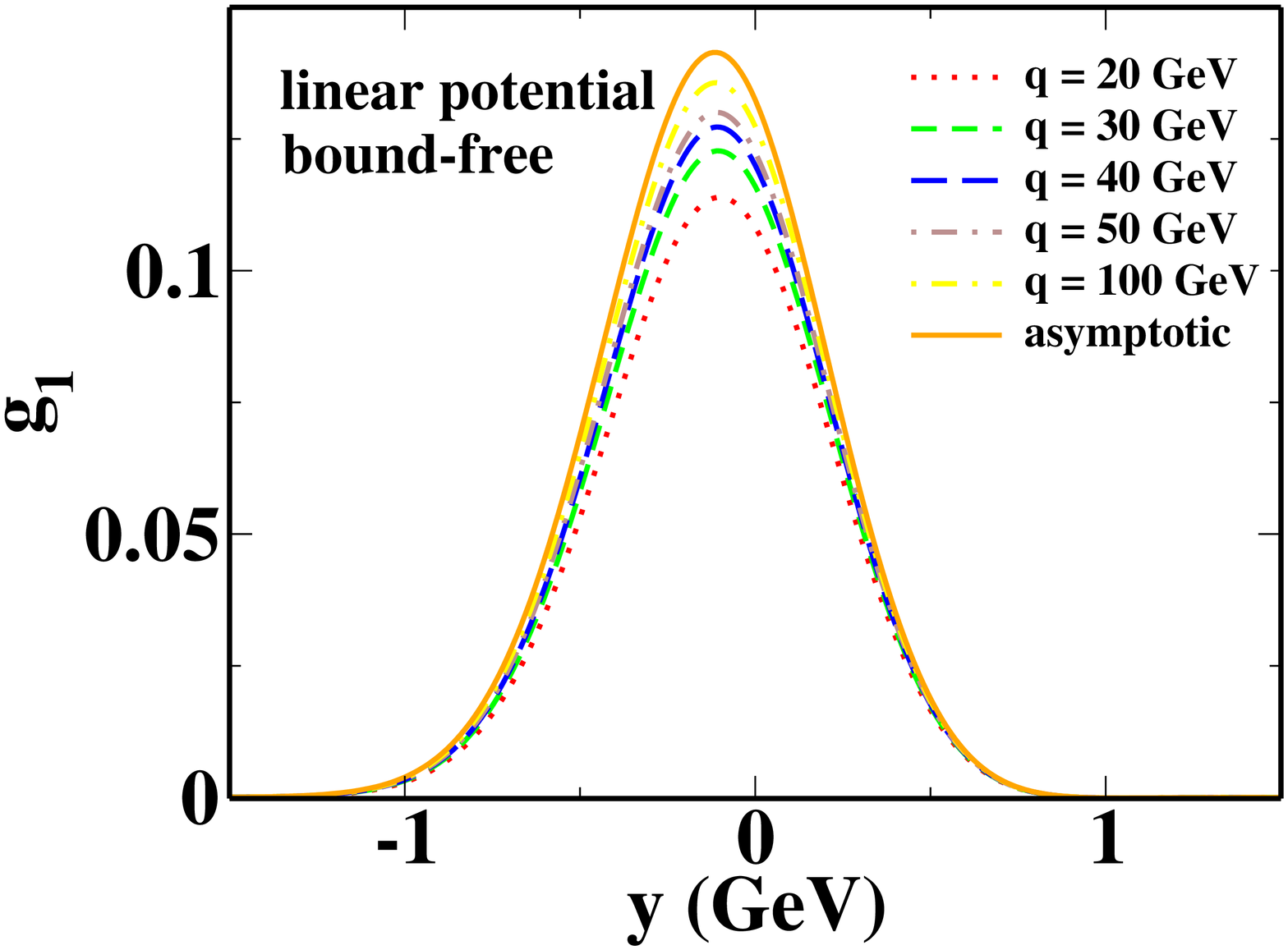}
\includegraphics[width=20pc,angle=270]{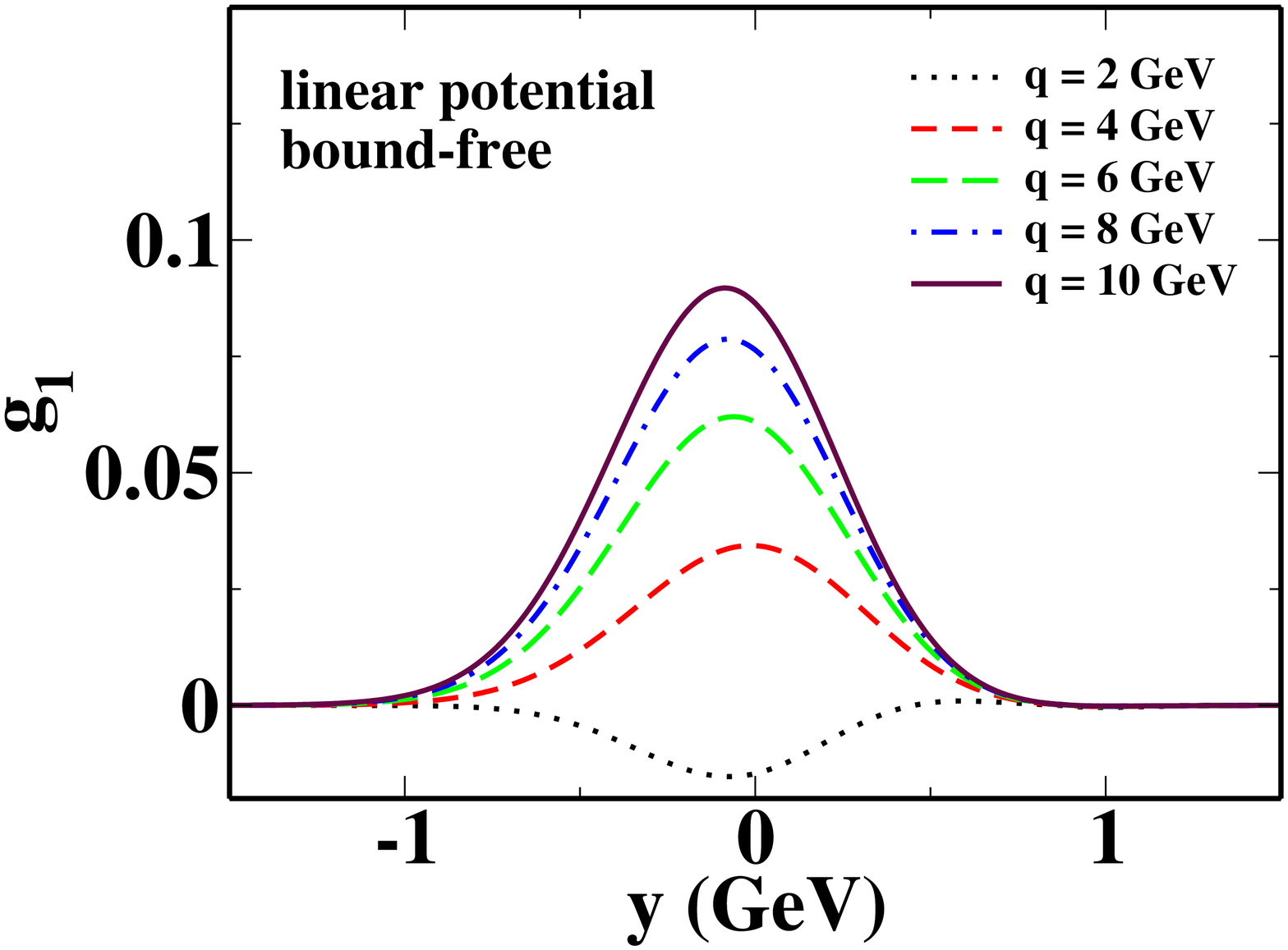}
\caption{The spin structure function $g_1$ is plotted
versus $y$ for several low (bottom panel) and high (top panel) values
of the three-momentum transfer $q$. The results shown have been
calculated for the bound-free transition, using the linear potential.}
\label{figbfg1}
\end{figure}

From this, we can see that the transverse-primed response $R_{T'}$
shows a scaling behavior very similar to $R_T$ (as discussed in
\cite{dirac}), and that the only interference response that is
accessible in inclusive electron scattering, $R_{TL'}$, exhibits
the earliest onset of scaling.

Now, we will consider the $y$-scaling behavior of the spin
structure functions. The spin structure functions are simply
combinations of the spin-dependent responses multiplied with some
kinematic factors, see Section \ref{secformalism}. We have already
seen how the relevant responses, $R_{T'}$ and $R_{TL'}$, scale.
Taking a look at the kinematic coefficients shows us that the
onset of $y$-scaling will be determined by the $q$ values at which
the kinematic factors reach their asymptotic values. For the
convenience of the reader, we recall Eq.~(\ref{defpolobs}) here:

\begin{equation}
\displaystyle{g_1 = - \frac{1}{2} M \frac{\nu^2}{q^2} \left (
R_{T'} + \frac{1}{\sqrt{2}} \frac{Q^2}{q \nu} R_{TL'} \right ) = -
\frac{1}{2} M \frac{\nu^2}{q^2} R_{T'} -  \frac{1}{2 \sqrt{2}} M
\frac{\nu Q^2}{ q^3}  R_{TL'}  } \nonumber
\end{equation}

\begin{equation}
\displaystyle{ g_2 = \frac{1}{2} M \frac{\nu^2}{q^2} \left( R_{T'}
- \frac{1}{\sqrt{2}} \frac{\nu}{q} R_{TL'} \right ) = \frac{1}{2}
M \frac{\nu^2}{q^2} R_{T'} - \frac{1}{2 \sqrt{2}} M
\frac{\nu^3}{q^3} R_{TL'}} \nonumber
\end{equation}

The asymptotic values of the kinematic factors multiplying the
responses are, for fixed $y$ and large $q$,

\begin{eqnarray}
\frac{\nu^2}{q^2} & \to & 1 + \frac{1}{q} 2 (y - E_0) \to 1 \nonumber\\
\frac{\nu Q^2}{q^3} & \to & \frac{1}{q} 2 (E_0 - y) \to 0 \nonumber\\
\frac{\nu^3}{q^3} & \to & 1 + \frac{1}{q} 3 (y - E_0) \to 1
\label{ssfkinfac}
\end{eqnarray}

leading to the asymptotic values of the spin structure functions:

\begin{eqnarray}
g_1 & \to & - \frac{1}{2} M R_{T'} \nonumber \\
g_2 & \to & \frac{1}{2} M (R_{T'} - \frac{1}{\sqrt{2}} R_{TL'}) \,
. \label{ssfasym}
\end{eqnarray}

\begin{figure}[ht]
\includegraphics[width=20pc,angle=270]{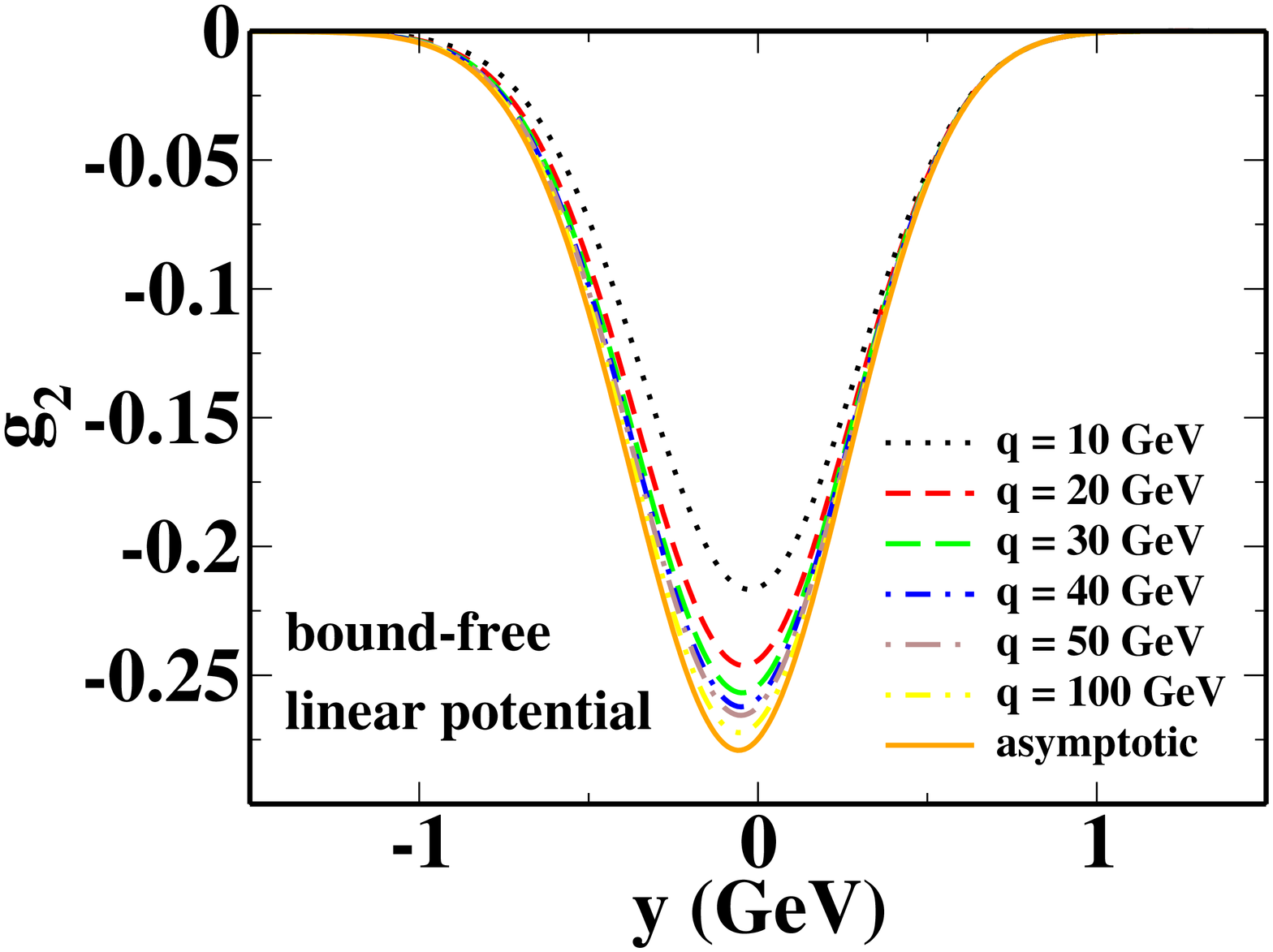}
\includegraphics[width=20pc,angle=270]{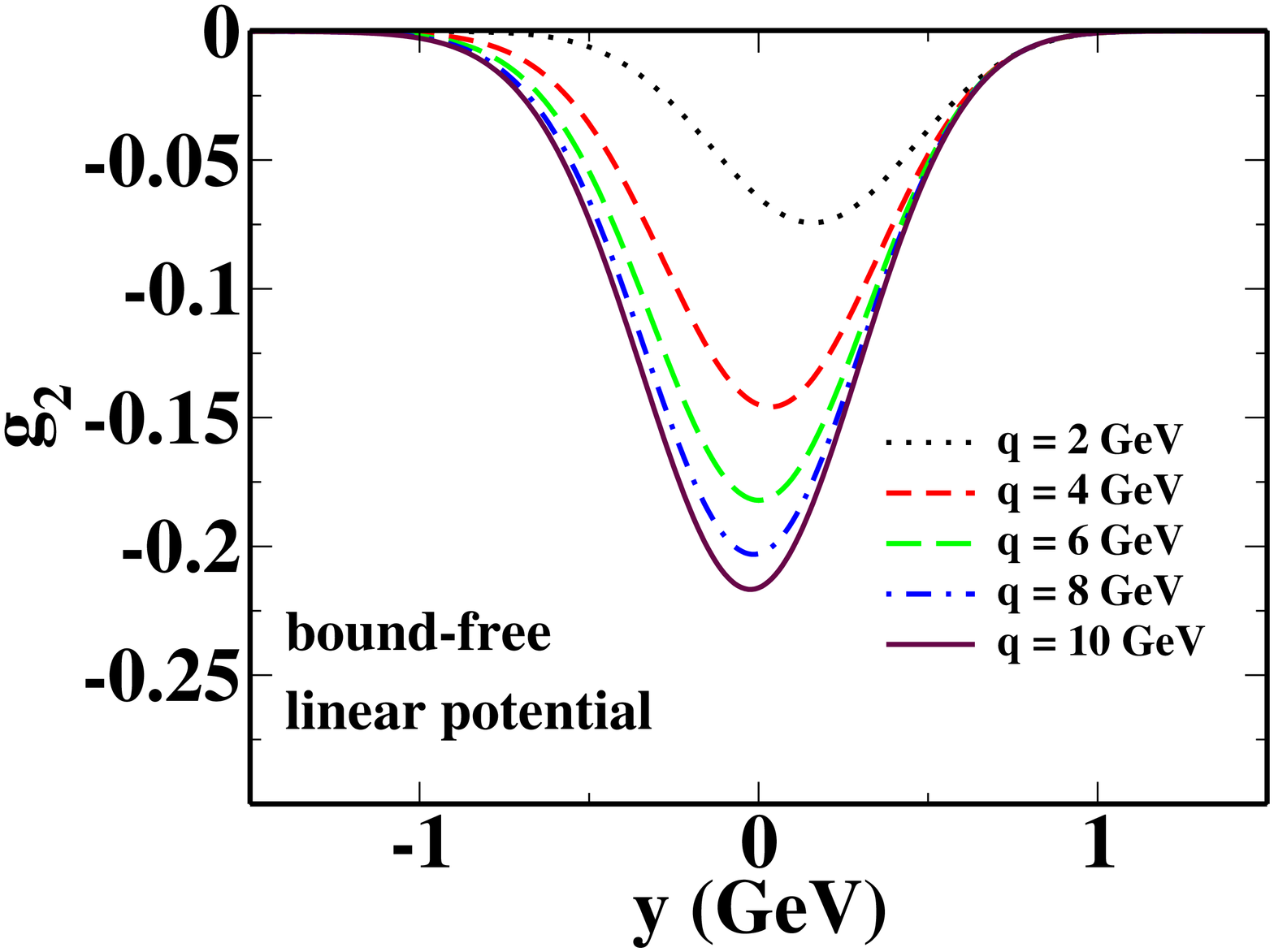}
\caption{The spin structure function $g_2$ is plotted
versus $y$ for several low (bottom panel) and high (top panel) values
of the three-momentum transfer $q$. The results shown have been
calculated for the bound-free transition, using the linear potential.}
\label{figbfg2}
\end{figure}

From Eq. (\ref{ssfkinfac}), one can see that the relevant scale
for the onset of scaling is given by the ground state energy
$E_0$. All responses peak roughly in the region $y \approx 0$, so
that the three-momenta $q$ necessary to reach the asymptotic
values are determined by $E_0$ alone. The spin structure function
$g_1$ for the bound-free transition is shown in
Fig.~\ref{figbfg1}.  The lower panel shows the spin structure
function for several $q$ values up to $10$~GeV.  One can see
clearly that $g_1$ changes significantly for each increase in $q$,
and that convergence has not set in at $q = 10 GeV$. This behavior
is due to the slow approach of the kinematic factors in $g_1$ to
their asymptotic values. The responses themselves scale much
faster, and are very close to their asymptotic values at $q =
10~GeV$, whereas the spin structure function $g_1$ is off its
asymptotic value by almost $50 \%$. Results for $g_2$ at lower
(bottom panel) and higher (top panel) three-momentum transfers are
shown in Fig.~\ref{figbfg2}. Just like $g_2$, it approaches its
scaling limits only at high momentum transfers.

\begin{figure}[ht]
\includegraphics[width=20pc,angle=270]{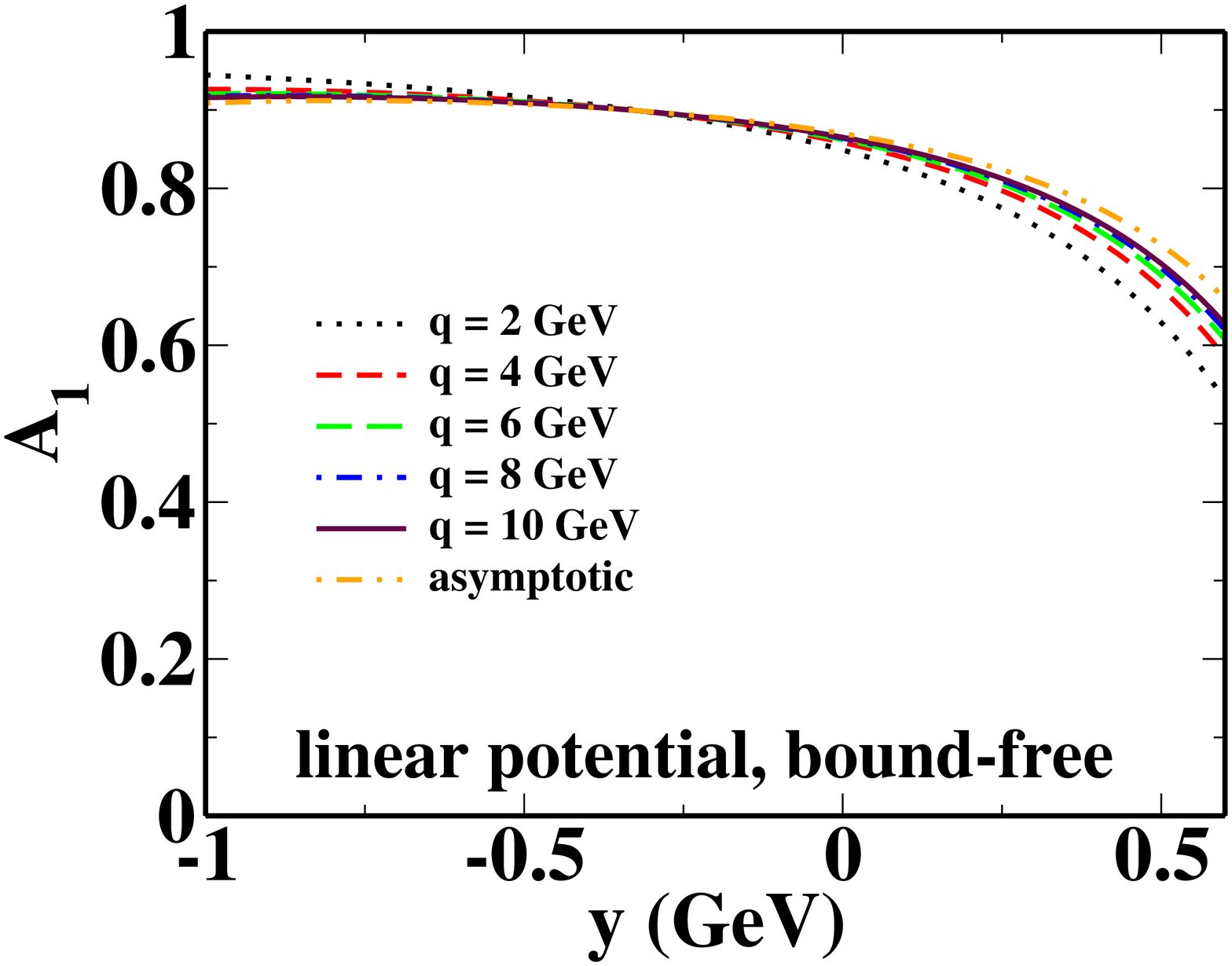}
\includegraphics[width=20pc,angle=270]{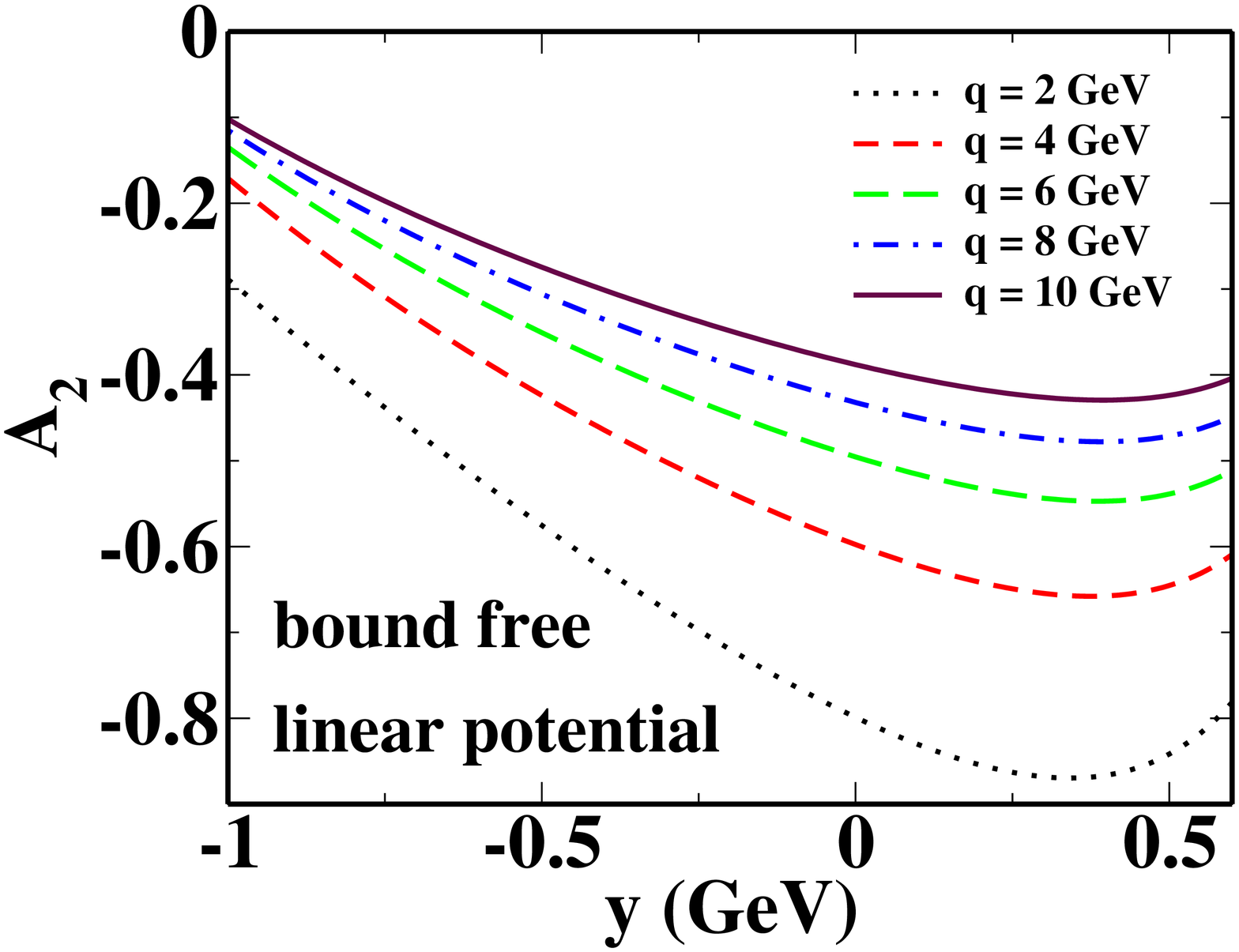}
\caption{The polarization asymmetries $A_1$ (top panel) and  $A_2$
(bottom panel) are plotted versus $y$ for several values of the
three-momentum transfer $q$, as well as their asymptotic value for $q
\to \infty$. The results shown have been calculated for the bound-free
transition, using the linear potential.}
\label{figbfa1a2}
\end{figure}

The polarization asymmetries $A_1$ and $A_2$ are ratios of response
functions, and may therefore show a different scaling behavior than
the responses themselves.  As we have seen that the two purely
transverse responses, $R_T$ and $R_{T'}$, have a very similar approach
to scaling, one may expect to see an even more rapid scaling of $A_1$,
the ratio of the two transverse responses.  The results for $A_1$ are
shown in the top panel of Fig.~\ref{figbfa1a2}. As $A_1$ is a pure
ratio of two responses that scale at reasonable values of $q$, without
any kinematic factors, we can expect $A_1$ to scale fast. Indeed, one
clearly sees that in the region of $ -0.5 < y < 0$, where the
responses are largest, $A_1$ scales almost immediately. Only the $q =
2~GeV$ curve differs very slightly from all the other curves,
including the asymptotic value, in this $y$ interval. Outside of that
region, where the responses have less strength, the scaling takes a
bit longer. In all cases, $q = 10 GeV$ is very close to the asymptotic
value.

The approach to scaling of the polarization asymmetry $A_2$ is
shown in the bottom panel of Fig.~\ref{figbfa1a2}.  Note that $A_2
\to 0$ for $q \to \infty$ and fixed $y$, because the kinematic
factor $\frac{Q}{q}$ goes to zero in this limit and the responses
scale in $y$. Thus, the approach to scaling of $A_2$ will mainly
be determined by the kinematic factor $\frac{Q}{q}$. So, even
though both $R_{TL'}$ and $R_T$ scale at reasonable values of $q$,
the scaling of $A_2$ is delayed. The differences between the
curves for increasing $q$ values decrease, but the result for $q =
10 GeV$ is still far away from the asymptotic value of $0$. Just
as for the spin structure functions, the kinematic factor that is
present determines the very slow onset of scaling.

In summary, we observe that the response functions reach values very
close to their asymptotic values around $q = 10~GeV$, and scale at the
latest at $q = 40~GeV$. The response $R_{TL'}$ is the fastest scaling
response, and the response $R_{T'}$ is the response that scales most
slowly. The spin structure functions and the polarization asymmetry
$A_2$ scale only for much higher momentum transfers, due to the
kinematic factors in their definitions. The polarization asymmetry
$A_1$, on the other hand, is the observable for which we observe
scaling at the lowest momentum transfers.

\begin{figure}[ht]
\includegraphics[width=20pc,angle=270]{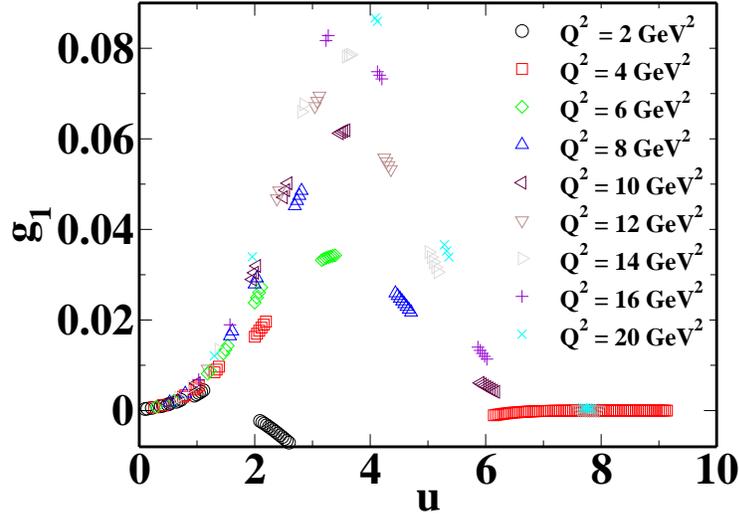}
\caption{The spin structure function $g_1$ is plotted versus $u$
for several values of the four-momentum transfer squared $Q^2$.
The results shown have been calculated for the bound-free
transition, using the linear potential.} \label{figuqsqg1}
\end{figure}

Note that the scaling behavior of the discussed observables does
not change when we use an ``x-type'' scaling variable, as in our
previous papers \cite{ijmvo,jvod2}. This is demonstrated in Fig.
~\ref{figuqsqg1}, where we show $g_1$ as a function of $u =
\frac{1}{2 m} (\sqrt{\nu^2 + Q^2} - \nu) (1 + \sqrt{1 + \frac{4
m^2}{Q^2}}) $, for fixed $Q^2$. Note that $u$ goes to $ u_{\infty}
= \frac{M_{targtet}}{m} x_{Bj}$ for large $Q^2$, its properties
have been discussed previously \cite{ijmvo,jvod2}. This is the
appropriate scaling variable for use with fixed $Q^2$ even at low
$Q^2$.

We have generated the plot by starting out with our results as
functions of $q$ and $y$, then calculated the corresponding values
of the four-momentum transfer $Q^2$ and the scaling variable $u$.
The results were then sorted into $Q^2$ bins. This is why there
are no continuous lines in the figure, just single data points.

One can see clearly that the scaling behavior of $g_1$ is
independent of the variables chosen: just as seen in
Fig.~\ref{figbfg1} for fixed $q$ and the variable $y$, at low
$Q^2$, $g_1$ starts out negative, then changes sign, and takes a
very long time to scale. Even between the points for $Q^2 = 16~
GeV^2$ and $Q^2 = 20~ GeV^2$, there is a significant difference.
Thus, we see that u-scaling is just as slow for $g_1$ (and $g_2$
and $A_2$) as $y$-scaling. The scaling behavior does not change at
all with the use of different scaling variables.

\subsection{Scaling and low $q$ duality in the bound-bound transition}
\label{secbb}

Now we proceed to discuss the behavior of the responses, spin
structure functions, and polarization asymmetries for final states
consisting solely of resonances. We refer to this case as the
``bound-bound'' transition. The wave functions for the excited
states have to be obtained by solving the Dirac equation
numerically.  Due to the involved nature of the numerics, the
highest accessible momentum transfer is $q = 10~GeV$. Work on
extending our calculations up to $30~GeV$ using a WKB
approximation will be reported elsewhere. Here, we focus on the
approach to scaling and on duality at low $q$.

\begin{figure}[ht]
\includegraphics[width=20pc,angle=270]{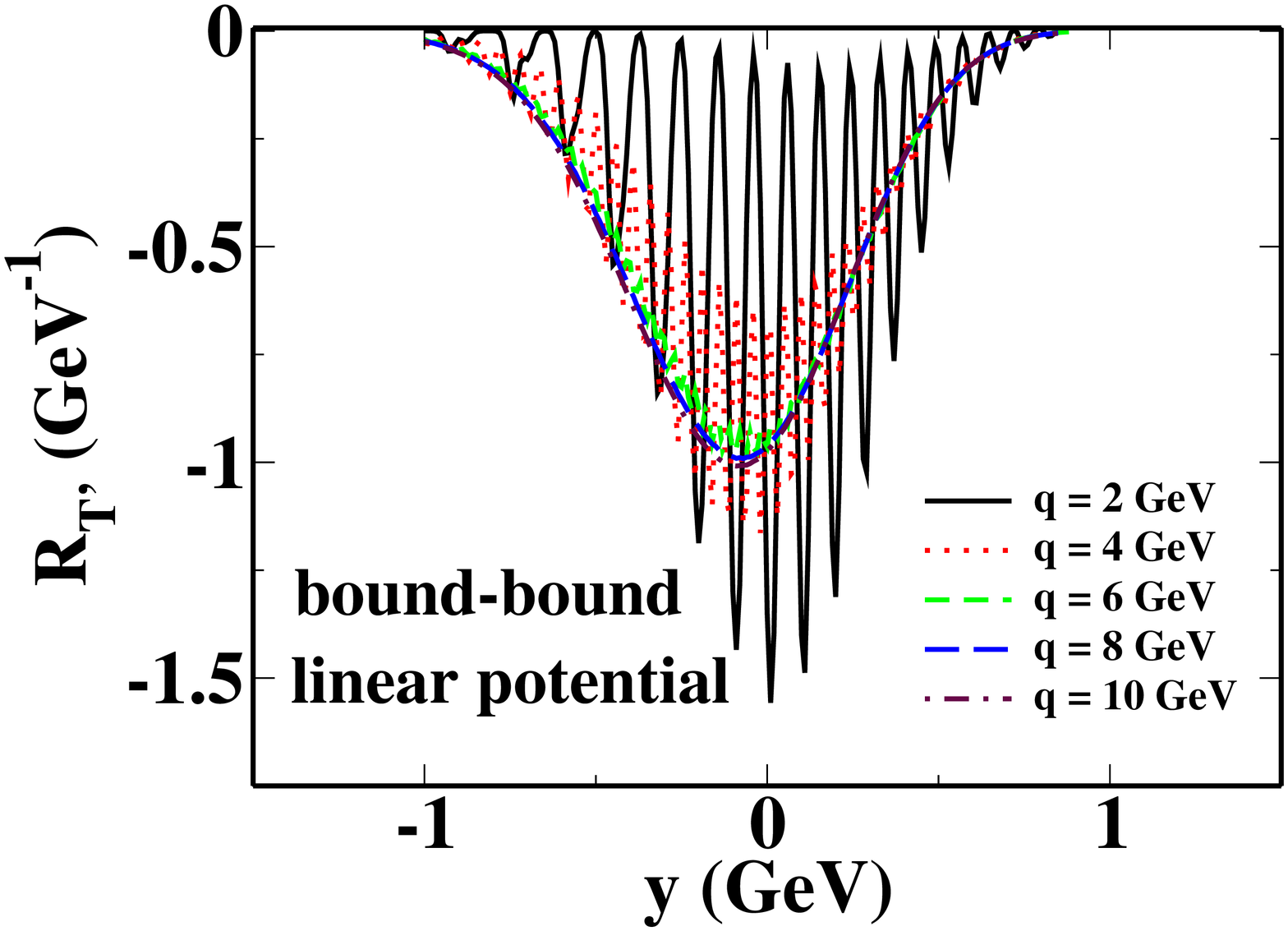}
\includegraphics[width=20pc,angle=270]{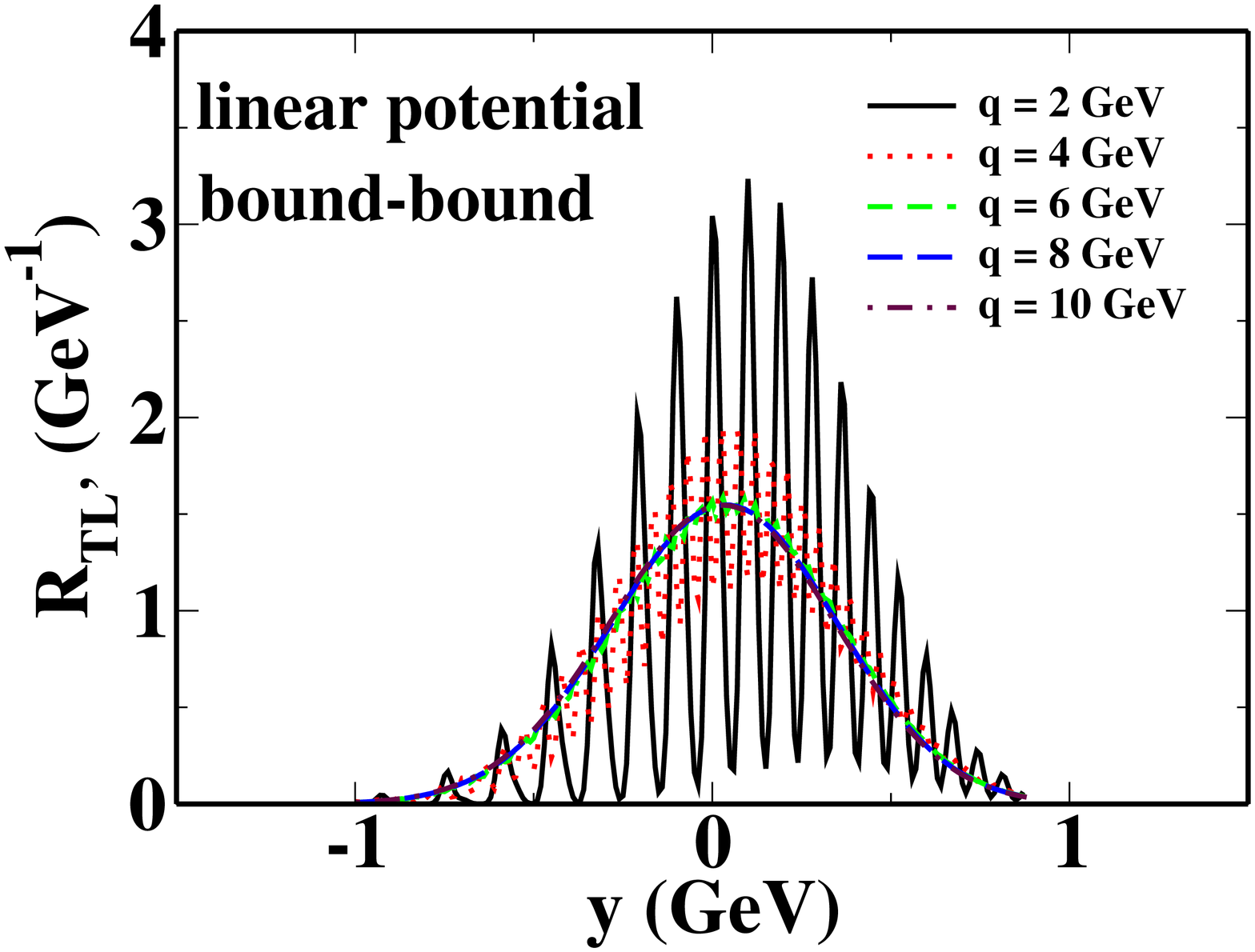}
\caption{The transverse-primed response function $R_{T'}$ (top panel)
and the transverse-longitudinal primed response function $R_{TL'}$
(bottom panel) are plotted versus $y$ for several values of the
three-momentum transfer $q$. The results shown have been calculated
for the bound-bound transition, using the linear potential.}
\label{figbbresp}
\end{figure}

We start again by considering the two polarized responses. We show
$R_{T'}$ in the top panel of Fig.~\ref{figbbresp}. The resonance
bumps at lower $q$ values give way to smooth curves formed by
many, closely spaced resonances. The graph shows that there is
still a small difference between the results for $q = 8~GeV$ and
$q = 10~GeV$, we have not yet reached the scaling value. This is
to be expected, as the bound-free transition scales only at higher
$q$ values for this response. In general, the more complicated
interplay of the various final state resonances leads to slower
scaling for the bound-bound transition. The resonances oscillate
around the smooth curves for higher momenta $q$ on the positive
$y$ flank, but are a bit below the smooth curves at the negative
$y$ flank. We see a very similar behavior for $R_{TL'}$, in the
bottom panel of Fig.~\ref{figbbresp}.

For $A_1$, the bound-bound transition results scale very quickly
in the region of the response peaks, $-0.5 < y < 0$. This is
consistent with the scaling behavior of $A_1$ as observed in the
bound-free transition, and stems from the definition of $A_1$ as
ratio of two fast-scaling responses. It is interesting to note
that the resonance bumps that are visible at $q = 2~GeV$ have
vanished almost completely at $q = 4~GeV$, even though they are
clearly present at $q = 4~GeV$ for the transverse and
transverse-primed responses, whose ratio forms $A_1$. This
exemplifies the very quick onset of scaling for this observable,
which we can observe even in the region of low momentum transfer
$q$.

Our current model gives at best a very qualitative insight on the
workings of duality for proton targets, as it is most closely
related to electron scattering off a heavy-light meson, i.e. a
B-meson. Nevertheless, our results seem to be very encouraging as
far as the prospects of applying quark-hadron duality to
extracting deep inelastic information on $A_1$ from data in the
resonance region are concerned.

\begin{figure}[ht]
\includegraphics[width=20pc,angle=270]{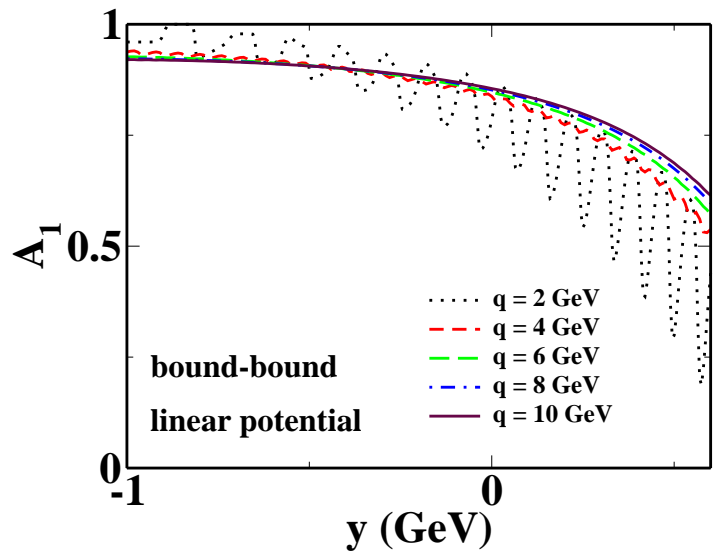}
\caption{The polarization asymmetry $A_1$ is plotted versus $y$
for several values of the three-momentum transfer $q$.  The
results shown have been calculated for the bound-bound transition,
using the linear potential.} \label{figbba1}
\end{figure}

\begin{figure}[ht]
\includegraphics[width=20pc,angle=270]{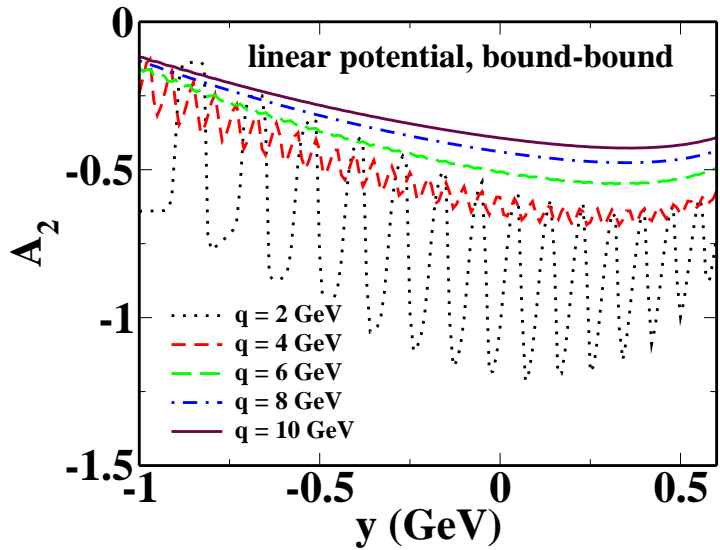}
\caption{The polarization asymmetry $A_2$ is plotted versus $y$
for several values of the three-momentum transfer $q$.  The
results shown have been calculated for the bound-bound transition,
using the linear potential.} \label{figbba2}
\end{figure}

The polarization asymmetry $A_2$, which contains a kinematic
factor multiplying the ratio of two responses, is shown in
Fig.~\ref{figbba2}. In contrast to $A_1$, $A_2$ changes
significantly with changing momentum transfer, and approaches its
scaling value of zero only slowly. As mentioned when discussing
the bound-free transition results for $A_2$ in the previous
section, the behavior of this polarization asymmetry is mainly
determined by the kinematic factor, not by the ratio of the
responses.

The same scenario - large variation for changes in momentum
transfer $q$ and slow approach to the scaling values - is repeated
for the spin structure functions $g_1$ and $g_2$.  As for $A_2$,
their behavior is determined by the slowly-scaling kinematic
factors, not by the fast-scaling responses. For the lowest
considered momentum transfer, $g_1$ even has a different sign,
compared to the higher momentum transfer results.

Our results suggest that the application of duality to resonance
region data for $A_1$ should be quite safe, while $A_2$, $g_1$,
and $g_2$ seem to be observables that are not amenable to such an
extraction procedure.

As far as duality in the spin structure function $g_1$ is
concerned, the available data
\cite{fatemi,amarian,hermesg1,slacg1} are not yet conclusive, and
may imply a different onset of duality for $g_1^p$ and $g_1^n$.
The Hermes data \cite{hermesg1} for $g_1^p$ indicate that the
onset of duality takes place at $Q^2 \approx 1.8 GeV^2$. A recent
analysis \cite{meziani} of Jefferson Lab data \cite{amarian} for
the first moment of $g_1^n$ indicates that duality could hold at
$Q^2$ as low as $1~GeV^2$. A dedicated Jefferson Lab experiment to
study duality in neutron spin structure functions in the resonance
region is currently being analyzed \cite{e01012}. Dong et al.
\cite{dongli,donghe,morechinese} recently used a constituent quark
model for the five lowest-lying resonances and various
parametrizations of scaling data for a careful theoretical
analysis of duality in $F_2$ and $g_1$ for the proton target. They
included target mass corrections and find that duality in $g_1$ is
not seen below $Q^2 \approx 2 GeV^2$. Close and Isgur
\cite{closeisgur}, using an $SU(6)$-symmetric, constituent quark
model, also predicted a slower onset of duality for $g_1$, and
pointed out that duality sets in faster in this model for $g_1^n$
than for $g_1^p$. Dominant magnetic interactions are necessary for
the symmetric quark model, and for the neutron target, this is
realized.

Our simple model is, of course, far from describing actual
electron scattering measurements. It is not intended to provide a
quantitative description of data, but to provide qualitative
insights into Nature. However, our findings on the scaling
behavior of the polarization asymmetries and spin structure
function might provide more general guidance: what we see is that
the onset of scaling is driven mainly by the kinematic factors
multiplying the reasonably fast-scaling response functions in the
definitions Eq.~(\ref{defpolobs}) of the spin structure functions
and polarization asymmetries. Where the factors are missing, i.e.
for $A_1$, convergence is rapid and the prospects for a successful
application of duality are very good. Where the factors are
present, and take a very high momentum to reach their asymptotic
value, e.g. in $g_1$, scaling is very slow and duality does not
hold. This suggests that the validity of quark-hadron duality in
spin observables may be strongly related to kinematic factors, and
less to dynamics. The faster scaling for $g_1^n$ might be
explained due to a smaller longitudinal-transverse interference
response $R_{TL'}$, see Eqs.~(\ref{ssfkinfac},\ref{ssfasym}).

However, one has to be cautious as certain features of nature are
not included in our model: it is well-known that the $\Delta$
resonance has an unnaturally small coefficient multiplying the
leading term and therefore gives rise to peculiarities in the
scaling and duality behavior, see \cite{carlnimay}.

\begin{figure}[ht]
\includegraphics[width=20pc,angle=270]{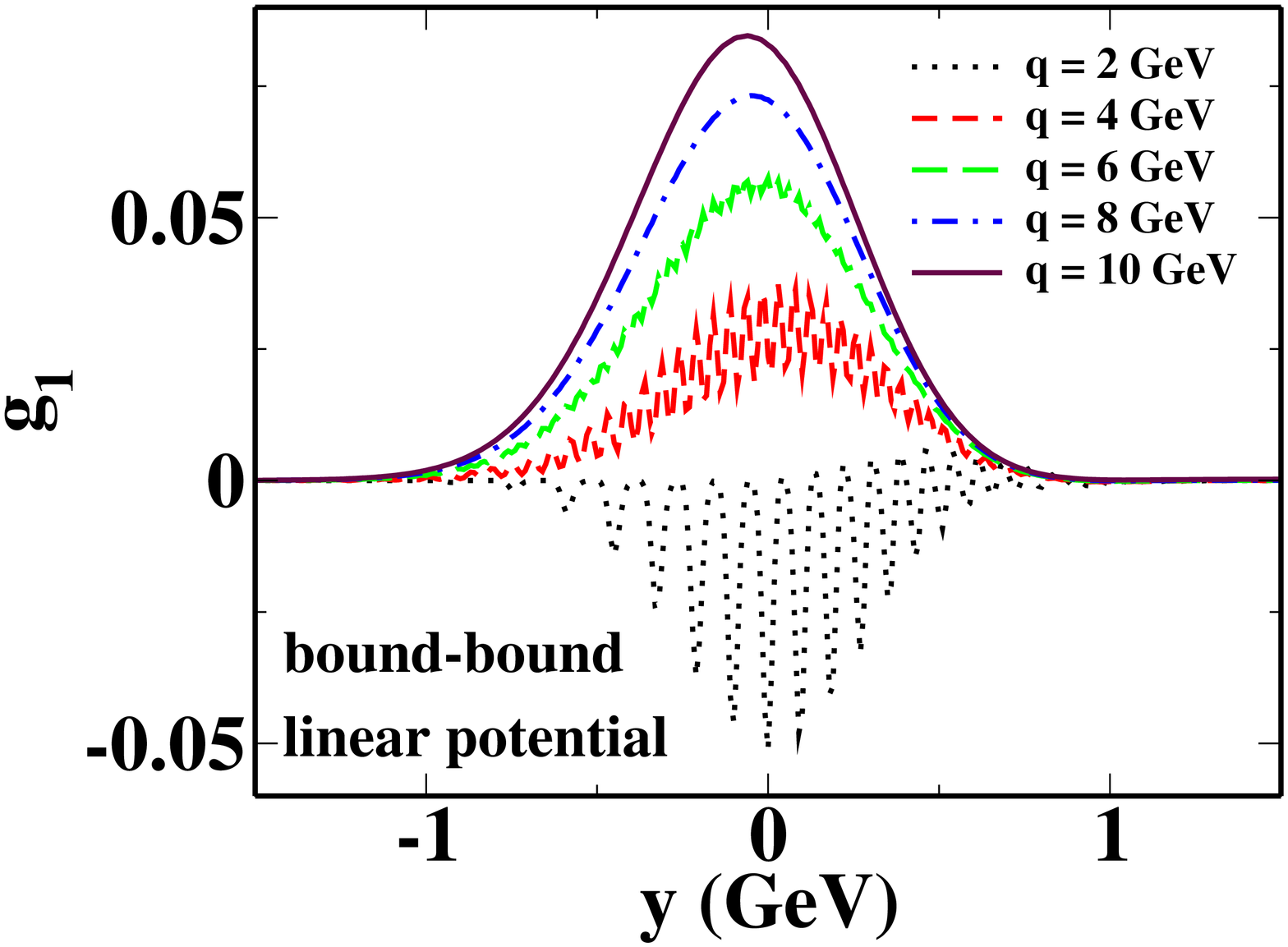}
\includegraphics[width=20pc,angle=270]{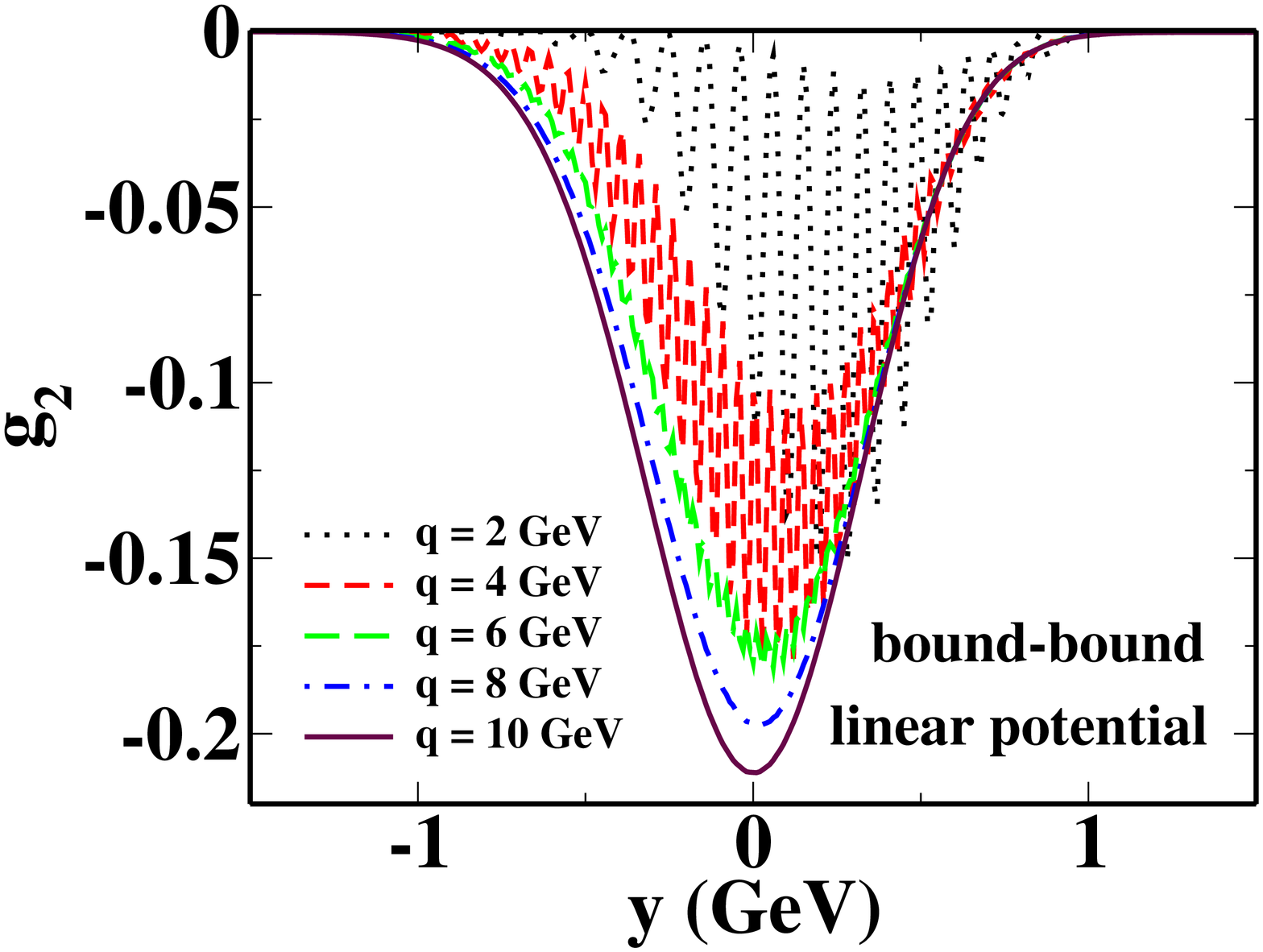}
\caption{The spin structure functions $g_1$ (top panel) and $g_2$
(bottom panel) are plotted versus $y$ for several values of the
three-momentum transfer $q$. The results shown have been
calculated for the bound-bound transition, using the linear
potential.} \label{figbbg1g2}
\end{figure}

\subsection{The role of the p-wave contribution}
\label{secpwave}

One interesting question is which role the p-wave in the ground
state plays. We will discuss the role of the ground state p-wave
contribution for the onset of scaling for the observables, and for
the scaling results themselves.

\subsubsection{The p-wave for the bound-free transition}

For the bound-free transition, we can read off from the momentum
distributions in Section \ref{secpwiayscal} that, without the
ground state p-wave,

\begin{equation}
n_v^0 (p) = n_s (p) = \frac{1}{2\pi} {\psi^{(+)}_{10\frac{1}{2}}}
^2(p) \, \, \mbox{and} \, \, n_v^s (p) = n_- (p) = 0 \,.
\end{equation}

\begin{figure}[ht]
\includegraphics[width=20pc,angle=270]{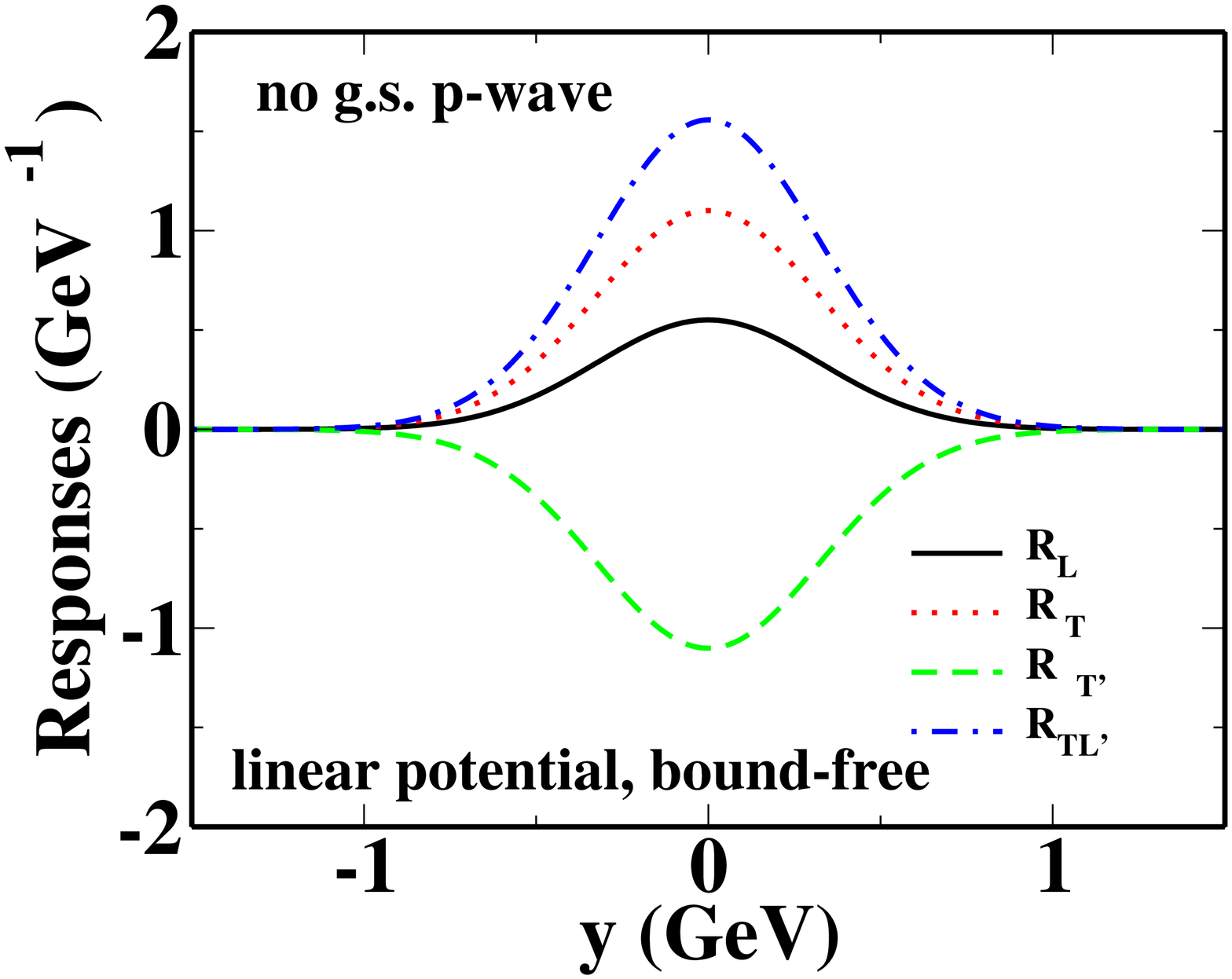}
\includegraphics[width=20pc,angle=270]{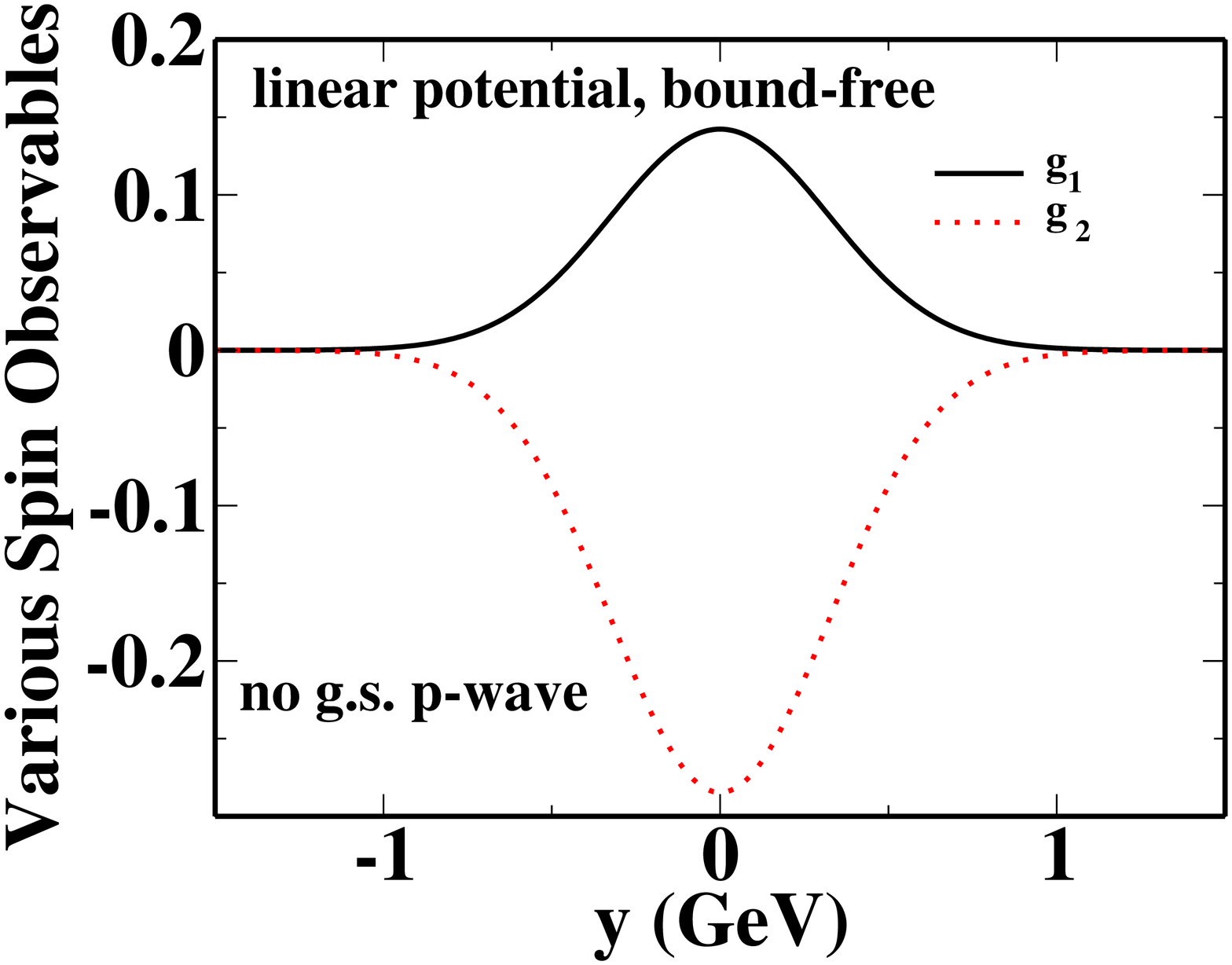}
\caption{The asymptotic behavior for $q \to \infty$ and fixed $y$
of the responses for the bound-free transition, without the
ground-state p-wave (top panel). We show $R_L$ (solid line), $R_T$
(dotted line), $R_{T'}$ (dashed line), and $R_{TL'}$ (dash-dotted
line). The asymptotic behavior for $q \to \infty$ and fixed $y$ of
the spin structure functions $g_1$ and $g_2$ for the bound-free
transition, without the ground-state p-wave (bottom panel). We
show $g_1$ (solid line) and $g_2$ (dotted line). Note that $A_2$
vanishes in this limit, and that $A_1 = 1$. } \label{figbfasymnop}
\end{figure}

Thus, in the scaling limit, we can introduce the response
\begin{equation}
R_{no \, p} = \frac{1}{16\pi^3}\int_{|y|}^{\infty}dp p
{\psi^{(+)}_{10\frac{1}{2}}}^2(p)
\end{equation}
and then write all the bound-free responses in the scaling limit
as
\begin{equation}
R_L = \frac{1}{2} R_{no \, p}, \, \, \, R_T = -R_{T'} = R_{no \,
p}, \, \, \, \, R_{TL'} = \sqrt{2} R_{no \, p} \,.
\end{equation}
From these equations, one sees that in the scaling limit, the peak
position is now identical for all four responses, and that the
responses differ only by a simple numerical factor. This leads
automatically to $A_1 = 1$ for all values of $q$ and $y$. Just
like the responses, the spin structure functions $g_1$ and $g_2$
also peak at $y = 0~GeV$ without the p-wave in the ground state.
Apart from slight shifts in the peak position, there are no major
changes in the asymptotic forms of the observables, see
Fig.~\ref{figbfasymnop}. For $R_L$, omission of the p-wave leads
to a slight reduction in peak height. $R_T$ also has a reduced
peak height (a reduction of about 10\%). In contrast, $R_{T'}$
maintains its peak height. The only response not to experience any
noticeable change is $R_{TL'}$, which was centered around $y =
0~GeV$ to start with, and does not change its peak height, either.

% make plots for all of the observables

The omission of the p-wave in the ground state simplifies the
analytic expressions obtained for the responses, and one expects
that scaling should set in more quickly. This is observed at low
$q$ values for $R_L$ and $R_T$. For $R_{T'}$ and $R_{TL'}$, the
presence of the p-wave contribution plays a negligible role in the
scaling behavior. This behavior is consistent with the fact that
the asymptotic forms of $R_L$ and $R_T$ are more strongly affected
by the omission of the p-wave.

\begin{figure}[ht]
\includegraphics[width=20pc,angle=270]{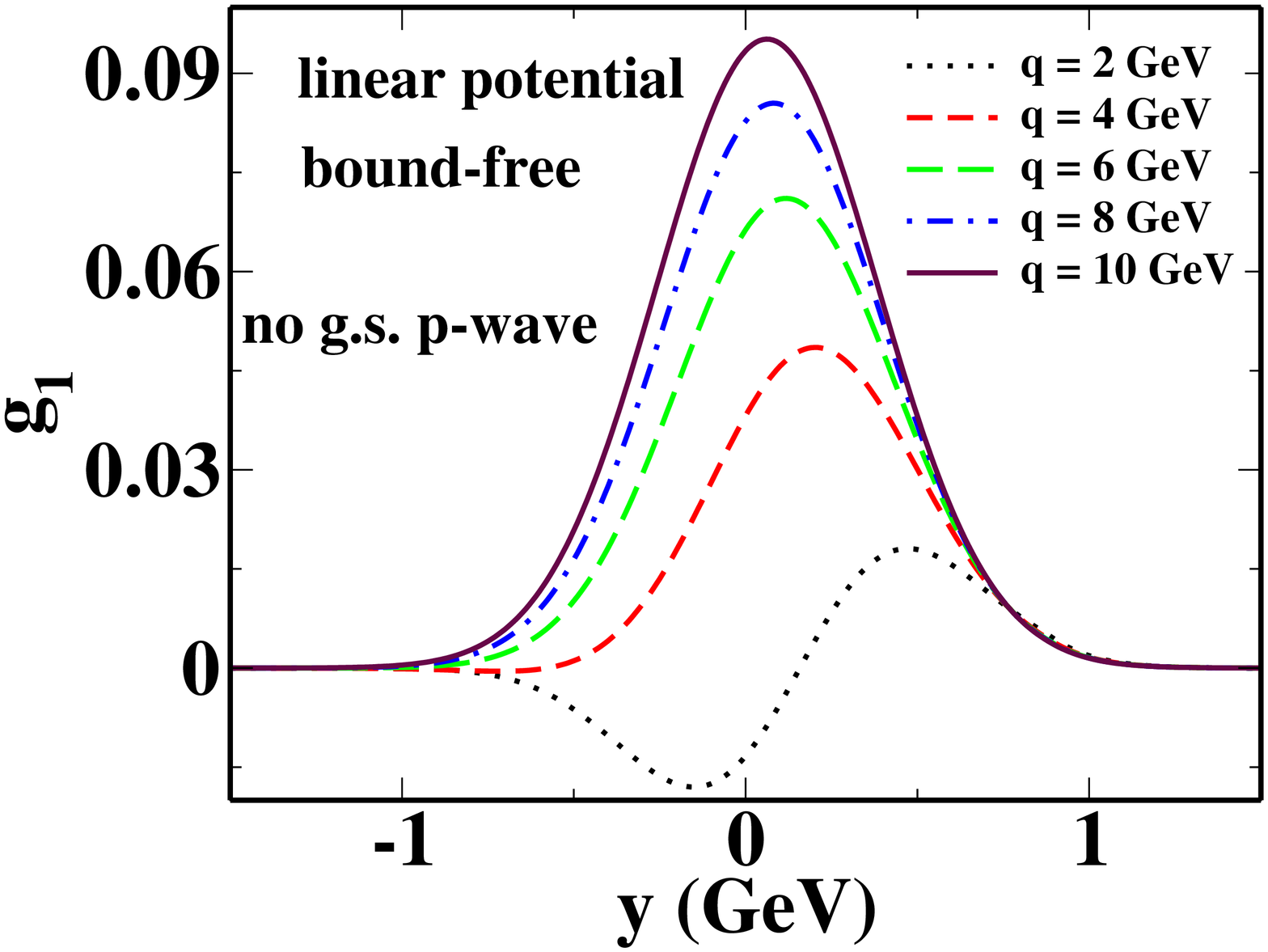}
\includegraphics[width=20pc,angle=270]{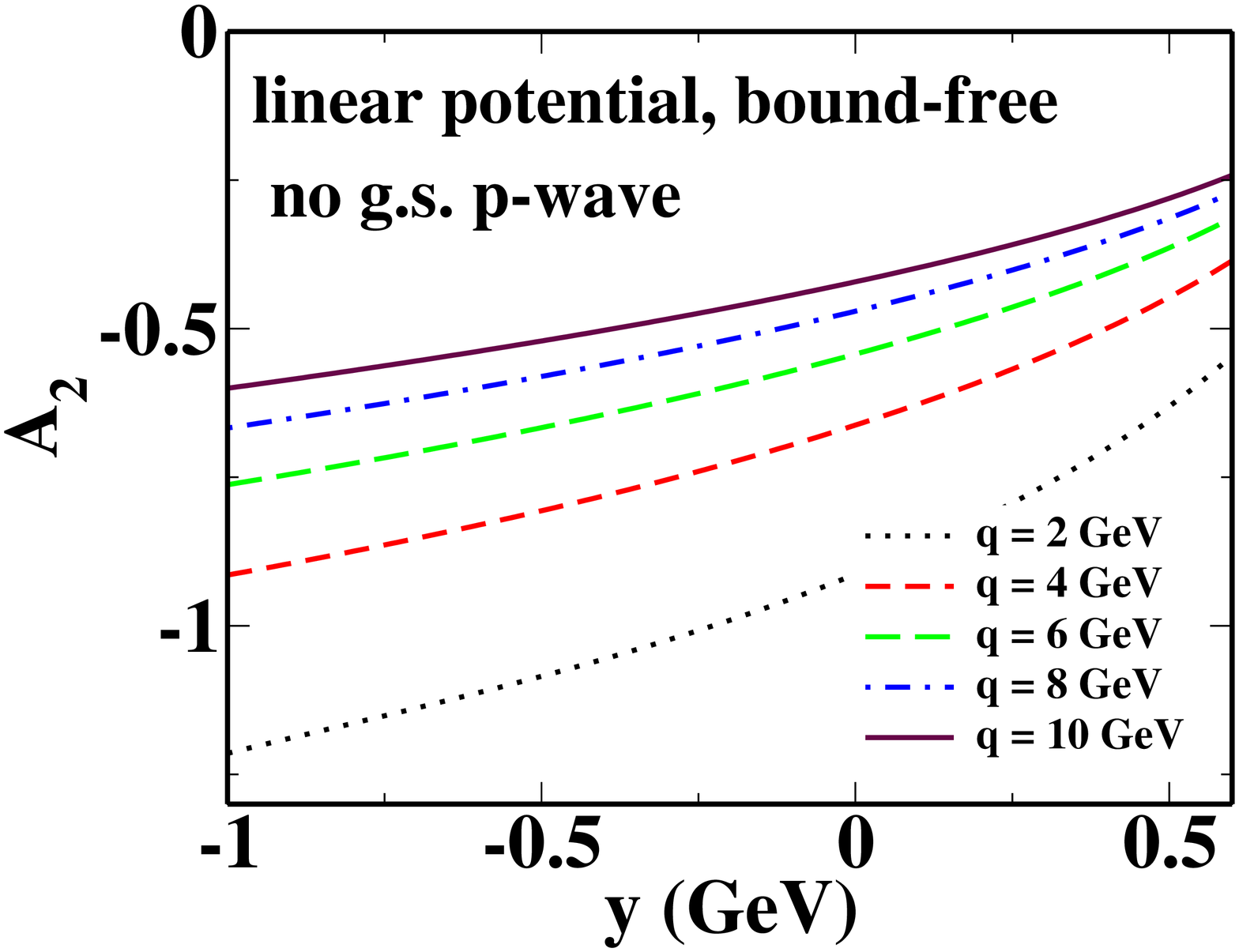}
\caption{The spin structure function $g_1$ (top panel) and the
polarization asymmetry $A_2$ (bottom panel) are shown without the
ground-state p-wave for several values of the three-momentum
transfer $q$ at fixed $y$.} \label{figbfg1a2nop}
\end{figure}

For $g_1$, the results change significantly for low $q$, see the
top panel of Fig.~\ref{figbfg1a2nop}. At $q = 2 ~GeV$, the spin
structure function is changing sign at $y \approx 0.1~GeV$ and
becomes positive. For higher $q$ values, $g_1$ peaks at slightly
higher $y$ values without p-wave, and the peak height is somewhat
higher, too. The onset of scaling is very slow, independent of the
p-wave contribution.

The onset of scaling is dramatically accelerated for $A_1$ by
dropping the p-wave, as $R_T$ and $R_T'$ coincide apart from an
overall sign, and $A_1$ is the ratio of these two responses.
Therefore, $A_1$ scales directly to $1$ for even the lowest value
of $q$, if there is no p-wave present.

While $A_2$ is still scaling slowly, it has quite a different
shape now, see the bottom panel of Fig.~\ref{figbfg1a2nop}.
Instead of starting at its maximum value at large negative $y$,
and decreasing to a minimum around $y \approx 0.4~GeV$, without
the p-wave, it starts out at its minimum value at large negative
$y$ and steadily increases.

\subsubsection{The p-wave for the bound-bound transition}

The situation for the bound-bound transition is similar: for the
longitudinal response, we observe small changes in peak position
and peak height due to the omission of the p-wave, just like for
the bound-free transition. For the transverse response, we see a
more pronounced reduction in peak height, as well as the already
familiar shift in the peak position. The transverse-longitudinal
primed response remains largely unaffected by the omission of the
p-wave, while the transverse primed response has a shifted peak
position, and no reduction in peak height. These observations are
very similar to the observations made above concerning the effects
of the p-wave on the asymptotic shapes of the responses for the
bound-free transition.

The polarization asymmetry $A_1$ again takes the value of $1$
immediately, as $R_T$ and $R_{T'}$ only differ by a sign. $A_2$
without the ground state p-wave has the same shape for bound-bound
and bound-free transitions, markedly different from the shape
including the p-wave. The spin structure function $g_1$ behaves
similarly for bound-bound and bound-free transitions, too. We do
not present any figures for the bound-bound results, as they are
so similar to the bound-free transition results shown above.

Strictly speaking, one would have to renormalize the remaining
wave function when switching off the p-wave. However, the effect
of this rescaling will be small, and we omit it here. The goal of
this discussion was to learn where the p-wave is important, and
where not. Summarizing, the p-wave has the biggest impact on the
polarization asymmetries, where switching off the p-wave leads to
immediate scaling for $A_1$ and a different shape for $A_2$,
before $A_2$ reaches its asymptotic value of zero. The spin
structure function $g_1$ also shows sensitivity to the p-wave
contribution at low $q$.

Our results for the role of the p-wave in $A_1$ confirm the
results in \cite{marka1}. There, the authors noted a distinct
suppression of $A_1$ at low Nachtmann $\xi$ due to the p-waves.
The region of large positive $y$, where we observe a fall-off of
$A_1$ when it is calculated with the complete wave function, and
find a value of $1$ when $A_1$ is calculated without the p-wave
contribution, corresponds to the low $\xi$ region.

\subsection{Model dependence}
\label{secpotmodels}

In this section, we discuss the influence of the chosen potential
on the results for the observables. For brevity, we restrict
ourselves to the bound-bound transition. Previously \cite{dirac},
we found that there is no qualitative difference between the
linear, static Coulomb, and running Coulomb potential results in
the longitudinal and transverse responses. Peak height and
position are different, but the approach to scaling and the low
$q$ oscillatory behavior are qualitatively the same.

First, we discuss the results for the Coulomb potential in the
bound-bound transition. The responses agree qualitatively with the
results for the linear potential, with similar shapes, peak
heights, peak locations, and scaling behavior. The same holds for
the spin structure functions $g_1$ and $g_2$. For the polarization
asymmetry $A_1$, the decrease from values close to $1$ is slightly
more pronounced for positive $y$ values, but the differences are
still small.

\begin{figure}[ht]
\includegraphics[width=20pc,angle=270]{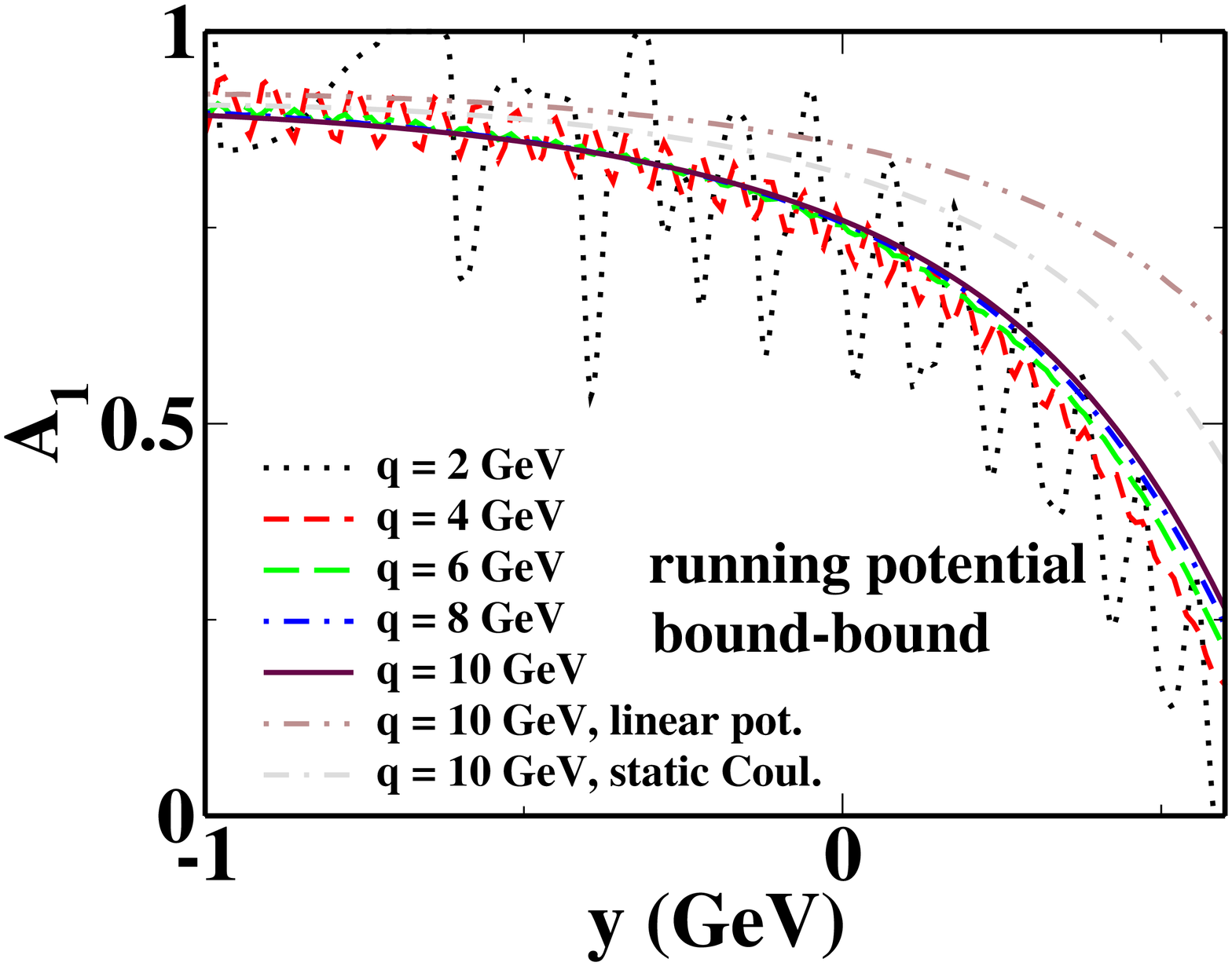}
\includegraphics[width=20pc,angle=270]{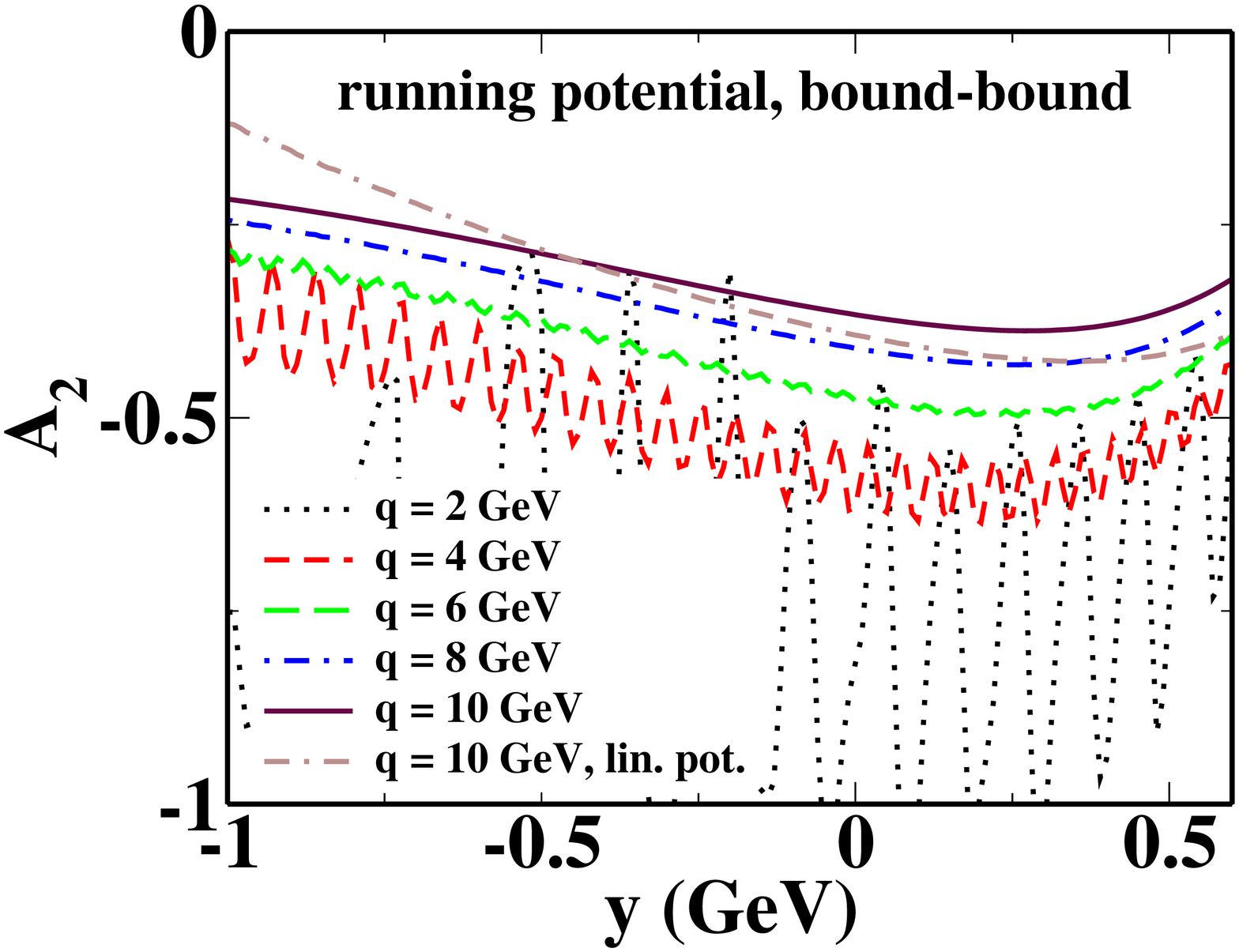}
\caption{The polarization asymmetries $A_1$ (top panel) and $A_2$
(bottom panel) are plotted versus $y$ for several values of the
three-momentum transfer $q$. The results shown have been
calculated for the bound-bound transition, using the running
Coulomb potential. The results for $q = 10~GeV$ for the linear
potential and the static Coulomb potential are also shown for
comparison.} \label{figa1run}
\end{figure}

This somewhat faster fall-off of $A_1$ is seen more distinctly for
the running potential, see Fig.~\ref{figa1run}. However, the
responses and spin structure functions look very similar for the
running Coulomb potential and the static Coulomb potential. $A_2$
has a slightly different shape for the running Coulomb potential,
it starts out more flatly at negative $y$ values than for the
linear potential and the static Coulomb potential. The onset of
convergence, however, does not seem to be affected noticeably by
the potential in the low $q$ region.

These results for the dependence of the spin observables on the
employed model potential are consistent with the results found for
the behavior of the longitudinal and transverse response
functions: while the results differ quantitatively, there is no
qualitative difference due to the different potentials, neither in
the results for the highest $q$ valuable attainable for
bound-bound transitions in our model, nor in the convergence of
the results.

The polarization asymmetries, as ratios of responses, are
naturally more sensitive than the other observables, but even they
only show a minor influence of the potential.

\section{Summary and Outlook}

We have presented results for responses, spin structure functions
and polarization asymmetries, calculated within a simple model
consisting of a light quark bound to an infinitely heavy
anti-quark (or a diquark) without charge. We have investigated the
onset of scaling and the scaling functions themselves in the
bound-free transition, where we are not hampered by the numerical
difficulties involved in calculating the bound-bound transition
for energies larger than $12~GeV$. For the bound-bound transition,
we looked at the approach to scaling and the duality at low
momentum transfer $q$.

We have found that the two polarization responses, $R_{T'}$ and
$R_{TL'}$, scale roughly like their unpolarized counterparts, the
longitudinal and transverse responses $R_L$ and $R_T$. The only
interference response, $R_{TL'}$, scales a bit faster than the
other responses, but the differences in scaling behavior are
small. Duality holds qualitatively for the responses, and for the
polarization asymmetry $A_1$. In fact, $A_1$ scales rapidly, and
judging from our model calculations, is the most promising
observable for the application of duality to extract DIS
information from resonance region data. These results are
reminiscent of the observation that the $A_1^p$ and $A_1^n$ data
show little $Q^2$ dependence \cite{anthony}.

In contrast, scaling sets in very slowly for $A_2$ and the spin
structure functions $g_1$ and $g_2$. This behavior is independent
of the type of scaling variable chosen: convergence for an x-type
scaling variable and large $Q^2$ is as slow as for the $y$-scaling
variable and increasing $q$. We have traced the reason for the
slow onset of scaling back to the kinematic factors multiplying
the responses in the expressions for the spin observables.

%Duality seems to be alive and kicking for the responses and $A_1$.
%The spin structure functions and $A_2$ seem to be less suitable
%for the application of duality, due to the slow scaling of the
%kinematic factors in their definitions.

Of course, these are results calculated within a simple model, and
the model should not be viewed as an attempt at a quantitative
description of electron scattering off a proton. In nature, it may
turn out that the scaling behavior of the responses is different,
and might compensate in some way for the slow scaling of the
kinematic factors. Still, our model calculations might provide
some useful guidance. It would be interesting to investigate if
the responses can be separated from the data available at present,
and to see if they indeed scale and exhibit duality as predicted
in our model.

As in \cite{dirac}, we have performed calculations for three
different potentials. The quantitative differences are small for
almost all observables, and qualitatively, a different potential
does not seem to change anything. The one exception is the
polarization asymmetry $A_1$, which shows a sensitivity to the
employed potential at high, positive values of $y$.

We have also studied the influence of the $p$-wave in the ground
state wave function on the observables. While most observables
undergo only small shifts in peak position and small changes in
peak height, our results do confirm the larger influence of the
$p$-wave contribution on $A_1$ at large positive $y$ values
observed in \cite{marka1}. However, the other observables seem to
be fairly robust with respect to the $p$-wave.

{\bf Acknowledgments}: The authors thank A. Radyushkin for a
discussion clarifying a point he made in a talk, and M. Paris for
discussions on his model calculations. S.J. thanks G. A. Miller
for a discussion on numerical methods. This work was supported in
part by funds provided by the U.S. Department of Energy (DOE)
under cooperative research agreement under No. DE-AC05-84ER40150
and by the National Science Foundation under grants No.
PHY-0139973 and PHY-0354916.


\begin{thebibliography}{299}

\bibitem{bgduality} E.~D.~Bloom and F.~J.~Gilman,
%``Scaling, Duality, and the Behavior of Resonances in Inelastic
%Electron-Proton Scattering''
Phys.\ Rev.\ Lett. {\bf 25}, 1140 (1970);
%``Scaling and the Behavior of Nucleon Resonances in
%Inelastic Electron-Nucleon Scattering'',
Phys.\ Rev.\ D {\bf 4}, 2901 (1971).


\bibitem{jlab}
I.~Niculescu et al.,
%``Evidence for valencelike quark hadron duality,''
Phys.\ Rev.\ Lett.\ {\bf 85}, 1182 (2000);
%``Experimental verification of quark hadron duality,''
Phys.\ Rev.\ Lett. {\bf 85}, 1186 (2000); R.~Ent, C.E.~Keppel and
I.~Niculescu,
%``Nucleon elastic form-factors and local duality,''
Phys. Rev. D {\bf 62}, 073008 (2000);  S.~Liuti, R.~Ent,
C.~E.~Keppel and I.~Niculescu,
%``Perturbative QCD analysis of local duality in a fixed W**2 framework,''
 Phys.\ Rev.\ Lett.\
{\bf 89}, 162001 (2002).

\bibitem{f1fldual}
Y.~Liang {\it et al.}  [Jefferson Lab Hall C E94-110
Collaboration],
%``Measurement of R = sigma(L)/sigma(T) and the separated longitudinal and
%transverse structure functions in the nucleon resonance region,''
arXiv:nucl-ex/0410027.


\bibitem{prdmom}
C.~S.~Armstrong, R.~Ent, C.~E.~Keppel, S.~Liuti, G.~Niculescu and
I.~Niculescu,
%``Moments of the proton F2 structure function at low Q**2,''
Phys.\ Rev.\ D {\bf 63}, 094008 (2001).


\bibitem{meziani}
Z.~E.~Meziani {\it et al.},
%``Higher twists and color polarizabilities in the neutron,''
arXiv:hep-ph/0404066.

\bibitem{emcdual}
J.~Arrington, R.~Ent, C.~E.~Keppel, J.~Mammei and I.~Niculescu,
%``Low-Q scaling, duality, and the EMC effect,''
arXiv:nucl-ex/0307012; J.~Arrington {\it et al.},
%``x and xi scaling of the nuclear structure function at large x,''
Phys.\ Rev.\ C {\bf 64}, 014602 (2001) [arXiv:nucl-ex/0102004].


\bibitem{hermes}
A.~Airapetian {\it et al.}  [HERMES Collaboration],
%``Evidence for quark-hadron duality in the proton spin asymmetry A1,''
Phys.\ Rev.\ Lett.\  {\bf 90}, 092002 (2003).

\bibitem{hermesg1}
A.~Fantoni  [HERMES Collaboration],
%``Quark hadron duality and Q**2 evolution of the GDH integral in the HERMES
%experiment,''
Eur.\ Phys.\ J.\ A {\bf 17}, 385 (2003).


\bibitem{e01012} Jefferson Lab experiment E01-012, J.-P. Chen, S.
Choi, and N. Liyanage, spokespersons.


\bibitem{12gevwp}
``The Science Driving the 12 GeV Upgrade of CEBAF'', White Paper,
Jefferson Lab 2001, edited by L. Cardman, R. Ent, N. Isgur, J.-M.
Laget, C. Leemann, C. Meyer, Z.-E. Meziani.




\bibitem{closeisgur}
F.~E.~Close and N.~Isgur,
%``The Origins of Quark-Hadron Duality:
%How Does the Square of the Sum Become the Sum of the Squares?'',
Phys. \ Lett.\ B {\bf 509}, 81 (2001).

\bibitem{closewallym}
F.~E.~Close and W.~Melnitchouk,
%``Symmetry breaking and quark hadron duality in structure functions,''
Phys.\ Rev.\ C {\bf 68}, 035210 (2003) [arXiv:hep-ph/0302013].

\bibitem{closezhaogpd}
F.~E.~Close and Q.~Zhao,
%``A pedagogic model for deeply virtual Compton scattering with quark-hadron duality,''
%Phys.\ Rev.\ D
{\bf 66}, 054001 (2002).

\bibitem{closezhao}
Q.~Zhao and F.~E.~Close,
%``Locality of quark hadron duality and deviations from quark counting  rules
%above resonance region,''
 Phys.\ Rev.\ Lett.\  {\bf 91}, 022004
(2003); Q.~Zhao and F.~E.~Close,
%``Restricted locality of quark-hadron duality in exclusive meson
%photoproduction reactions above the resonance region,''
arXiv:hep-ph/0411257.




\bibitem{carlnew}
C.~E.~Carlson,
%``What we know about the theoretical foundation of duality in electron
%scattering,''
arXiv:hep-ph/0005169; A.~Afanasev, C.~E.~Carlson and C.~Wahlquist,
%``Scaling and duality in semi-exclusive processes,''
Phys.\ Rev.\ D {\bf 62}, 074011 (2000)

\bibitem{carlnimay}
C.~E.~Carlson and N.~C.~Mukhopadhyay,
%``Bloom-Gilman duality in the resonance spin structure functions,''
Phys.\ Rev.\ D {\bf 58}, 094029 (1998); C.~E.~Carlson and
N.~C.~Mukhopadhyay,
%``Duality in the inelastic structure function and the disappearing Delta
%resonance,''
Phys.\ Rev.\ D {\bf 47}, 1737 (1993); C.~E.~Carlson and
N.~C.~Mukhopadhyay,
%``A Quantum Chromodynamic Explanation Of Bloom-Gilman Duality,''
Phys.\ Rev.\ D {\bf 41}, 2343 (1990).

\bibitem{donghe}
Y.~B.~Dong and J.~He,
%``Study Of Quark Hadron Duality Based On A Constituent Quark Model,''
Nucl.\ Phys.\ A {\bf 720}, 174 (2003).

\bibitem{dongli} Y.~B.~Dong and M.~F.~Li,
%``Nucleon spin structure function and quark-hadron duality,''
Phys.\ Rev.\ C {\bf 68}, 015207 (2003).

\bibitem{morechinese}
Y.~B.~Dong,
%``Study of target mass effects on the moments of g1 and Bloom-Gilman
%quark-hadron duality,''
Nucl.\ Phys.\ A {\bf 744}, 293 (2004); Y.~B.~Dong and J.~Liu,
%``Resonance contribution to nucleon structure functions and quark-hadron
%duality,''
Nucl.\ Phys.\ A {\bf 739}, 166 (2004); Y.~B.~Dong and Q.~G.~Feng,
%``Nucleon spin structure functions in the resonance region and the duality,''
Commun.\ Theor.\ Phys.\  {\bf 39}, 675 (2003); Y.~B.~Dong and
M.~F.~Li,
%``Nucleon structure function F2 in the resonance region and quark hadron
%duality,''
Commun.\ Theor.\ Phys.\  {\bf 39}, 193 (2003).


\bibitem{pp}
M. W. Paris and V. R. Pandharipande,
%``Scaling of space and
%timelike response of confined relativistic particles'',
Phys. \ Lett.\ B {\bf 514}, 361 (2001); M.~W.~Paris,
%``Spacelike And Timelike Response Of Confined Relativistic Particles,''
Eur.\ Phys.\ J.\ A {\bf 17}, 401 (2003); M.~W.~Paris and
V.~R.~Pandharipande,
%``Final state interaction contribution to the response of confined
%relativistic particles,''
Phys.\ Rev.\ C {\bf 65}, 035203 (2002).

\bibitem{mpdirac}
M.~W.~Paris,
%``Electromagnetic response of confined Dirac particles,''
Phys.\ Rev.\ C {\bf 68}, 025201 (2003).

\bibitem{marka1}
V.~R.~Pandharipande, M.~W.~Paris and I.~Sick,
%``Virtual photon asymmetry for confined, interacting Dirac particles with spin
%symmetry,''
arXiv:nucl-th/0410093; V.~R.~Pandharipande, M.~W.~Paris and
I.~Sick,
%``Spin asymmetries for confined Dirac particles,''
arXiv:nucl-th/0308078.



\bibitem{kiev}
R.~Fiore, A.~Flachi, L.~L.~Jenkovszky, A.~I.~Lengyel and
V.~K.~Magas,
%``Explicit model realizing parton hadron duality,''
Eur.\ Phys.\ J.\ A {\bf 15}, 505 (2002) [arXiv:hep-ph/0206027];
L.~Jenkovszky, V.~K.~Magas and E.~Predazzi,
%``Resonance-reggeon and parton-hadron duality in strong interactions,''
Eur.\ Phys.\ J.\ A {\bf 12}, 361 (2001) [arXiv:hep-ph/0110374].


\bibitem{leyaouancped}
A.~Le Yaouanc, D.~Melikhov, V.~Morenas, L.~Oliver, O.~Pene and
J.~C.~Raynal,
%``Duality in the non-relativistic harmonic
%oscillator quark model in the  Shifman-Voloshin limit: A
%pedagogical example,''
Phys.\ Lett.\ B {\bf 488}, 153 (2000).




\bibitem{simula}
G.~Ricco, M.~Anghinolfi, M.~Ripani, S.~Simula and M.~Taiuti,
%``Bloom-Gilman duality of inelastic structure functions in nucleon
%and  nuclei,''
Phys.\ Rev.\ C {\bf 57}, 356 (1998); S.~Simula,
%``Target-mass corrections and the Bloom-Gilman duality of the
%nucleon  structure function,''
Phys.\ Lett.\ B {\bf 481}, 14 (2000).




\bibitem{wallymprl}
W.~Melnitchouk,
%``Local Duality Predictions for x ~ 1 Structure Functions,''
Phys.\ Rev.\ Lett.\  {\bf 86}, 35 (2001) [Erratum-ibid.\  {\bf
93}, 199901 (2004)] [arXiv:hep-ph/0106073].


\bibitem{adelaide}
W.~Melnitchouk, K.~Tsushima and A.~W.~Thomas,
%``Quark-hadron duality and the nuclear EMC effect,''
Eur.\ Phys.\ J.\ A {\bf 14}, 105 (2002) [arXiv:nucl-th/0110071];
F.~M.~Steffens and K.~Tsushima,
%``Local duality and charge symmetry violation in quark distributions,''
Phys.\ Rev.\ D {\bf 70}, 094040 (2004) [arXiv:hep-ph/0408018];
K.~Tsushima, K.~Saito and F.~M.~Steffens,
%``Effect of bound nucleon internal structure change on nuclear structure
%functions,''
arXiv:hep-ph/0409217.

\bibitem{ijmvo}
N. Isgur, S. Jeschonnek, W. Melnitchouk, and J. W. Van Orden,
%``Quark-hadron duality in structure functions'',
Phys.\ Rev.\ D {\bf 64}, 054005 (2001).

\bibitem{jvod2}
S.~Jeschonnek and J.~W.~Van Orden,
%``Quark hadron duality in a relativistic, confining model,''
Phys.\ Rev.\ D {\bf 65}, 094038 (2002).

\bibitem{dirac}
S.~Jeschonnek and J.~W.~Van Orden,
%``Modeling quark hadron duality for relativistic, confined fermions,''
Phys.\ Rev.\ D {\bf 69}, 054006 (2004) [arXiv:hep-ph/0310298].


\bibitem{elba}
J.~W.~Van Orden and S.~Jeschonnek,
%``Energy-weighted Sum Rules, y-scaling and Duality'',
Eur. Phys. J. {\bf A17}, 391 (2003).

\bibitem{myhugs}
M.~A.~DeWitt and S.~Jeschonnek,
%``Quark-hadron duality: A pedagogical introduction,''
%\href{http://www.slac.stanford.edu/spires/find/hep/www?irn=5980003}{SPIRES entry}
{\it Prepared for 17th Annual HUGS at CEBAF (HUGS 2002), Newport
News, Virginia, 3-21 Jun 2002}.

\bibitem{zoltanfranz}
Z.~Batiz and F.~Gross,
%``Quark hadron duality and scaling in reduced QCD,''
Phys.\ Rev.\ D {\bf 69}, 074006 (2004) [arXiv:nucl-th/0310088].


\bibitem{liuti}
S.~Liuti,
%``Violations of parton hadron duality in deep inelastic scattering,''
Eur.\ Phys.\ J.\ A {\bf 17}, 397 (2003); N.~Bianchi, A.~Fantoni
and S.~Liuti,
%``Parton hadron duality in unpolarised and polarised structure functions,''
Phys.\ Rev.\ D {\bf 69}, 014505 (2004) [arXiv:hep-ph/0308057].

\bibitem{davidovsky}
V.~V.~Davidovsky and B.~V.~Struminsky,
%``Nucleon structure functions, resonance form-factors, and duality,''
Phys.\ Atom.\ Nucl.\  {\bf 66}, 1328 (2003) [Yad.\ Fiz.\  {\bf
66}, 1368 (2003)]; V.~V.~Davidovsky and B.~V.~Struminsky,
%``The behavior of form factors of nucleon resonances and quark hadron
%duality,''
arXiv:hep-ph/0205130.

\bibitem{hofmann}
R.~Hofmann,
%``Operator product expansion and local quark-hadron duality: Facts and
%riddles,''
Prog.\ Part.\ Nucl.\ Phys.\  {\bf 52}, 299 (2004)
[arXiv:hep-ph/0312130]; R.~Hofmann,
%``Towards OPE based local quark-hadron duality: Light-quark channels,''
Nucl.\ Phys.\ B {\bf 623}, 301 (2002) [arXiv:hep-ph/0109008].


\bibitem{richural}
R.~F.~Lebed and N.~G.~Uraltsev,
%``Precision studies of duality in the 't Hooft model,''
Phys.\ Rev.\ D {\bf 62}, 094011 (2000).

\bibitem{semilep}
See e.g. I.~Bigi and N.~Uraltsev,
%``A vademecum on quark hadron duality,''
hep-ph/0106346;I.~Bigi, M.~Shifman, N.~Uraltsev and A.~Vainshtein,
%``Heavy flavor decays, OPE and duality in two-dimensional 't Hooft model,''
Phys.\ Rev.\ D {\bf 59}, 054011 (1999); B.~Grinstein and
R.~F.~Lebed,
%``Quark hadron duality in the 't Hooft model for meson weak decays:  Different quark diagram
%topologies,''
Phys.\ Rev.\ D {\bf 59}, 054022 (1999); B.~Grinstein and
R.~F.~Lebed,
%``Explicit quark-hadron duality in heavy-light
%meson weak decays in the  't Hooft model,''
Phys.\ Rev.\ D {\bf 57}, 1366 (1998); C.~G.~Boyd, B.~Grinstein and
A.~V.~Manohar,
%``Semileptonic B and $\Lambda_b$ Decays and Local Duality in QCD,''
 Phys.\ Rev.\ D {\bf 54}, 2081 (1996).



\bibitem{isgurwise}
N.~Isgur and M.~B. Wise,
%``Excited charm mesons in semileptonic
%anti-B decay and their contributions to a Bjorken sum rule,''
Phys.\ Rev.\ D {\bf 43}, 819 (1991).

\bibitem{ralf}
see e.g. R. Rapp and J. Wambach, {\sl Adv. Nucl. Phys.} {\bf 25},
1 (2000).


\bibitem{qcdsr}
M.~A.~Shifman, A.~I.~Vainshtein and V.~I.~Zakharov,
%``QCD And Resonance Physics: Applications,''
Nucl.\ Phys.\ B {\bf 147}, 448 (1979); M.~A.~Shifman,
A.~I.~Vainshtein and V.~I.~Zakharov,
%``QCD And Resonance Physics. Sum Rules,''
 Nucl.\ Phys.\ B {\bf 147}, 385
(1979); A.~I.~Vainshtein, V.~I.~Zakharov, V.~A.~Novikov and
M.~A.~Shifman,
%``Asymptotic Freedom In Quantum Mechanics,''
 Sov.\
J.\ Nucl.\ Phys.\  {\bf 32}, 840 (1980); E.~C.~Poggio, H.~R.~Quinn
and S.~Weinberg,
%``Smearing The Quark Model,''
Phys.\ Rev.\ D {\bf 13}, 1958 (1976);
 A.V.~Radyushkin,
%``Introduction to QCD Sum Rule Approach,''
in {\em Strong Interactions at Low and Intermediate Energies}, ed.
J.L.~Goity (World Scientific, 2000), [hep-ph/0101227];
T.~D.~Cohen, R.~J.~Furnstahl, D.~K.~Griegel and X.~Jin,
%``QCD sum rules and applications to nuclear physics,''
Prog.\ Part.\ Nucl.\ Phys.\ {\bf 35}, 221 (1995).

\bibitem{minerva}
D.~Drakoulakos {\it et al.}  [Minerva Collaboration],
%``Proposal to perform a high-statistics neutrino scattering experiment using a
%fine-grained detector in the NuMI beam,''
arXiv:hep-ex/0405002.

\bibitem{Horowitz:2004yf}
C.~J.~Horowitz, M.~A.~Perez-Garcia and J.~Piekarewicz,
%``Neutrino-pasta scattering: The opacity of nonuniform neutron-rich matter,''
Phys.\ Rev.\ C {\bf 69}, 045804 (2004) [arXiv:astro-ph/0401079].

\bibitem{vipuli}
Vipuli Dharmawarda, private communication.

\bibitem{diehl}
M.~Diehl,
%``Generalized parton distributions,''
Phys.\ Rept.\  {\bf 388}, 41 (2003) [arXiv:hep-ph/0307382].

\bibitem{haiyan}
H.~Gao and L.~Zhu,
%``Pion photoproduction on the nucleon,''
arXiv:nucl-ex/0411014.


\bibitem{dgp}
A.De~R\'ujula, H.~Georgi and H.D.~Politzer,
%``Demythification Of Electroproduction, Local Duality And Precocious Scaling,''
Annals
Phys.\  {\bf 103}, 315 (1977);
%``An Explanation Of Local Duality And Precocious Scaling,''
Phys.\ Lett.\ B {\bf 64}, 428 (1977).



\bibitem{jiunrau}
X.~Ji and P.~Unrau,
%``Parton - hadron duality: Resonances and
%higher twists,''
 Phys.\ Rev.\ D {\bf 52}, 72 (1995);
%
X.~Ji and P.~Unrau,
%``Q**2 dependence of the proton's G1 structure
%function sum rule,''
 Phys.\ Lett.\ B {\bf 333}, 228 (1994);
%
X.~Ji and W.~Melnitchouk,
%``Spin-dependent twist-four matrix
%elements from g1 data in the resonance  region,''
Phys.\ Rev.\ D {\bf 56}, 1 (1997).

\bibitem{nathana1n}
N. Isgur,
%``Valence Quark Spin Distribution Functions'',
Phys.\ Rev.\ D {\bf 59}, 034013 (1999).

\bibitem{a1nprc}
X.~Zheng {\it et al.}  [Jefferson Lab Hall A Collaboration],
%``Precision measurement of the neutron spin asymmetries and spin-dependent
%structure functions in the valence quark region,''
arXiv:nucl-ex/0405006.

\bibitem{waka}
M.~Wakamatsu,
%``Light-flavor sea-quark distributions in the nucleon in the SU(3) chiral
%quark soliton model. I. Phenomenological predictions,''
Phys.\ Rev.\ D {\bf 67}, 034005 (2003); M.~Wakamatsu,
%``Light-flavor sea-quark distributions in the nucleon in the SU(3) chiral
%quark soliton model. II: Theoretical formalism,''
Phys.\ Rev.\ D {\bf 67}, 034006 (2003).

\bibitem{a1datajlab}
Xiachao Zheng {\it et al.},
%``Precision measurement of the neutron spin asymmetry $A_1^n$ and spin-flavor decomposition in the
%valence quark region'',
Phys.\ Rev.\ Lett.\ {\bf 92}, 012004 (2004).


\bibitem{smca1n}
D.~Adams {\it et al.}  [Spin Muon Collaboration],
%``A New measurement of the spin dependent structure function g1(x) of the
%deuteron,''
Phys.\ Lett.\ B {\bf 357}, 248 (1995).

\bibitem{hermesa1n}
K.~Ackerstaff {\it et al.}  [HERMES Collaboration],
%``Measurement of the neutron spin structure function g1(n) with a  polarized
%He-3 internal target,''
Phys.\ Lett.\ B {\bf 404}, 383 (1997) [arXiv:hep-ex/9703005].

\bibitem{slacg1}
K.~Abe {\it et al.}  [E143 Collaboration],
%``Measurement of the proton and deuteron spin structure function g1 in  the
%resonance region,''
Phys.\ Rev.\ Lett.\  {\bf 78}, 815 (1997); K.~Abe {\it et al.}
[E154 Collaboration],
%``Precision determination of the neutron spin structure function g1(n),''
Phys.\ Rev.\ Lett.\  {\bf 79}, 26 (1997) [arXiv:hep-ex/9705012].

\bibitem{slaca1n}
P.~L.~Anthony {\it et al.}  [E142 Collaboration],
%``Deep inelastic scattering of polarized electrons by polarized He-3 and  the
%study of the neutron spin structure,''
Phys.\ Rev.\ D {\bf 54}, 6620 (1996) [arXiv:hep-ex/9610007].

\bibitem{nilanga}
N. Liyanage, private communication.

\bibitem{waw}
J.~Zeng, J.~W.~Van Orden and W.~Roberts,
%``Heavy mesons in a relativistic model,''
Phys.\ Rev.\ D {\bf 52}, 5229 (1995).

\bibitem{godfreyisgur}
S.~Godfrey and N.~Isgur,
%``Mesons In A Relativized Quark Model With Chromodynamics,''
Phys.\ Rev.\ D {\bf 32}, 189 (1985).

\bibitem{donnellyras}
T.~W.~Donnelly and A.~S.~Raskin,
%``Considerations Of Polarization In Inclusive Electron Scattering From
%Nuclei,''
Annals Phys.\  {\bf 169}, 247 (1986).

\bibitem{weithom}
A.~W.~Thomas and W.~Weise, {\sl The Structure of the Nucleon},
Wiley-Vch, Germany (2001).

\bibitem{jifil}
B.~W.~Filippone and X.~D.~Ji,
%``The spin structure of the nucleon,''
Adv.\ Nucl.\ Phys.\  {\bf 26}, 1 (2001) [arXiv:hep-ph/0101224].

\bibitem{fatemi}
R.~Fatemi {\it et al.}  [CLAS Collaboration],
%``Measurement of the proton spin structure function g1(x,Q**2) for Q**2  from
%0.15-GeV**2 to 1.6-GeV**2 with CLAS,''
Phys.\ Rev.\ Lett.\  {\bf 91}, 222002 (2003).

\bibitem{amarian}
M.~Amarian {\it et al.},
%``The Q**2 evolution of the generalized Gerasimov-Drell-Hearn integral  for
%the neutron using a He-3 target,''
Phys.\ Rev.\ Lett.\  {\bf 89}, 242301 (2002).

\bibitem{anthony}
P.~L.~Anthony {\it et al.}  [E155 Collaboration],
%``Measurements of the Q**2 dependence of the proton and neutron spin
%structure functions g1(p) and g1(n),''
Phys.\ Lett.\ B {\bf 493}, 19 (2000) [arXiv:hep-ph/0007248].

\end{thebibliography}
\end{document}